\documentclass[preprint,12pt]{aastex}

\newcommand{\kms}{km~s$^{-1}$}

\newcommand{\ldl}{$\lambda/{\Delta}{\lambda}$}
\newcommand{\teff}{T$_{eff}$}

\newcommand{\meth}{CH$_4$}
\newcommand{\water}{H$_2$O}
\newcommand{\name}{2MASS~J15200224$-$4422419}
\newcommand{\namesh}{2MASS~J1520$-$4422}

\slugcomment{Submitted to ApJ 6 Sep 2006; accepted 21 Nov 2006}

\shorttitle{2MASS J1520-4424AB}
\shortauthors{Burgasser et al.}

\begin{document}

\title{Discovery of a High Proper Motion L Dwarf Binary: 2MASS~J15200224$-$4422419AB }

\author{Adam J.\ Burgasser\altaffilmark{1,2},
Dagny L.\ Looper\altaffilmark{2,3},
J.\ Davy Kirkpatrick\altaffilmark{4},
and
Michael C.\ Liu\altaffilmark{3}}

\altaffiltext{1}{Massachusetts Institute of Technology, Kavli Institute for Astrophysics and Space Research,
Building 37, Room 664B, 77 Massachusetts Avenue, Cambridge, MA 02139; ajb@mit.edu}
\altaffiltext{2}{Visiting Astronomer at the Infrared Telescope Facility, which is operated by
the University of Hawaii under Cooperative Agreement NCC 5-538 with the National Aeronautics
and Space Administration, Office of Space Science, Planetary Astronomy Program.}
\altaffiltext{3}{Institute for Astronomy, University of Hawaii, 2680 Woodlawn Drive, Honolulu, HI 96822}
\altaffiltext{4}{Infrared Processing and Analysis Center, M/S
100-22, California Institute of Technology, Pasadena, CA 91125}

\begin{abstract}
We report the discovery of the wide
L1.5+L4.5 binary 2MASS J15200224-4422419AB,
identified during spectroscopic followup of high proper motion
sources selected from the Two Micron All Sky Survey.
This source was independently identified by Kendall et al. in
the SuperCOSMOS Sky Survey.
Resolved $JHK$ photometry and low resolution 
near-infrared spectroscopy demonstrate that this system is
composed of two
well-separated (1$\farcs$174$\pm$0$\farcs$016) L dwarfs.  Component
classifications are derived using both spectral ratios and 
comparison to near-infrared spectra of previously classified field L dwarfs.
Physical association for the pair is deduced from the large
($\mu$ = 0$\farcs$73$\pm$0$\farcs$03 yr$^{-1}$) 
common proper motion of the components and their similar 
spectrophotometric distances (19$\pm$2~pc).  The 
projected separation of the binary, 22$\pm$2~AU, is consistent with
maximum separation/total system mass trends for very low mass binaries.
The 2MASS J1520$-$4422 system exhibits both large tangential (66$\pm$7~{\kms})
and radial velocities ($-70{\pm}18$~{\kms}), and its motion in the
local standard of rest suggests that it is an old member of the
Galactic disk population.
This system joins a growing list of well-separated ($>$0$\farcs$5),
very low mass binaries, and is an excellent target 
for resolved optical spectroscopy to constrain its age as well as
trace activity/rotation trends near the hydrogen-burning limit.
\end{abstract}

\keywords{binaries: visual ---
stars: individual ({\name}) ---
stars: low mass, brown dwarfs
}

\section{Introduction}

Multiple stellar systems are important laboratories for 
a wide range of astrophysical phenomena,
from the formation and evolution of planetary systems
to the distribution of dark matter in the Galaxy.  Stellar multiples
drive novae outbursts in cataclysmic
variable systems; are useful
distance ladders for star clusters
in the local Group; and provide one of the few direct
means of measuring stellar mass.  The production of multiples
is inherent to the star formation process itself, and 
the aggregate properties of multiple systems in a given population
can provide insight into the genesis of that population.

Multiples are of particular importance in the study 
of very low mass (VLM; M $\leq$ 0.1M$_{\sun}$)
stars and brown dwarfs.
Several of the first brown dwarfs identified were found as companions
to nearby stars \citep{bec88,nak95,reb98}, and $>$75 binaries 
composed entirely of VLM dwarfs are now known 
\citep[and references therein]{me06ppv}.\footnote{A current list is
maintained by N.\ Siegler at \url{http://paperclip.as.arizona.edu/$\sim$nsiegler/VLM\_binaries}.}
The salient properties of these systems 
--- their small separations, high mass ratios and
low frequency --- have been used as empirical tests of brown dwarf
formation theories \citep{rei01,me03b,me06ppv,clo03,bou04,sie05}.
Astrometric monitoring of resolved doubles have provided direct measures
of mass for a few VLM binaries \citep{lan01,bou04,bra04,zap04},
while the recently identified eclipsing binary 2MASS J0535218-054608AB
has enabled the first radius measurement in the substellar regime
\citep{sta06}.  Coupled with independent age determinations, 
these observations
also constrain brown dwarf evolutionary theories \citep{zap04,sta06}.
Multiplicity provides explanations for other observed phenomena, such as peculiar features in individual brown dwarf
spectra \citep{cru04,mcg05,me0423} and unusual 
photometric trends across the transition
between L-type and T-type brown dwarfs
\citep{me06hst,liu06}.

Despite their fecundity, only a limited number of VLM binaries 
have resolved spectroscopic measurements.\footnote{Resolved 
spectroscopy of VLM binaries is reported in \citet{got02,pot02,bou04,cha04,luh04,mcc04,bil05,me06d2200,clo06,jay06,ken06,mce06,mar06}; and \citet{moh06}.}
Such observations provide detailed information on the 
components of these systems, and can 
clarify empirical
trends as a function of spectral type, temperature or mass
by eliminating the variables of age and composition
(assuming coeval formation).
One regime in which such conditions can be exploited 
is the transition between the M
and L dwarf spectral classes \citep{kir05}.  Occurring at an effective temperature 
{\teff} $\approx$ 2300~K \citep{gol04,vrb04}, 
this transition is characterized by the formation of photospheric
condensates that dominate the near-infrared spectra of 
L dwarfs \citep{tsu96a,jon97,bur99,all01,ack01};
a rapid decline in the frequency and strength of H$\alpha$
emission, related to the presence of a hot chromosphere
\citep{giz00,moh02,wes04}; and an
increase in mean rotational velocities, possibly related to
a change in angular momentum loss mechanisms or age effects \citep{moh03a}.
In addition,
late M and early L spectral types
encompass the transition between hydrogen burning stars and 
brown dwarfs, depending on the age of the source \citep{cha00,bur01}. 

Our group is currently conducting a near-infrared proper motion survey using
multi-epoch data from 
the Two Micron All Sky Survey \citep[hereafter 2MASS]{skr06}.
The goal of this program is to identify late-type, 
nearby and/or high velocity
sources that may have been missed by existing photographic plate proper motion
surveys,
or near-infrared searches based solely on color-selection criteria.
Details on the survey will
be presented in a forthcoming publication (J.\ D.\ Kirkpatrick et al.,
in preparation).  
Here we report the identification of a well-resolved
L dwarf binary system, {\name} (hereafter {\namesh}).
This source has been independently identified by 
\citet{ken06} by combining 2MASS and SuperCOSMOS Sky Survey
\citep[hereafter SSS]{ham01a,ham01b,ham01c} catalog data.
Our initial identification of this source is described in $\S$~2.
Its recognition as a binary in follow-up imaging observations,
and measurements of separation, position angle and flux ratios
are described in $\S$~3.
In $\S$~4 we present near-infrared spectroscopic
observations obtained with the SpeX spectrograph \citep{ray03}
mounted on the 3m NASA Infrared Telescope Facility (IRTF).
These include resolved low-resolution spectra that identify
the components as L dwarfs, and composite moderate-resolution
data that reveal detailed spectral features similar to previously 
studied L dwarfs.
Analysis of our observations is provided in $\S$~5, 
including spectral classification of the components, 
kinematics of the system and additional binary parameters.
We discuss {\namesh} in the context of other well-resolved binaries
in $\S$~6, and motivate future spectroscopic investigation
to apply the ``binary Li test''.
Results are summarized in $\S$~7.

\section{Identification of {\namesh}}

\begin{figure}
\epsscale{1.0}
\plotone{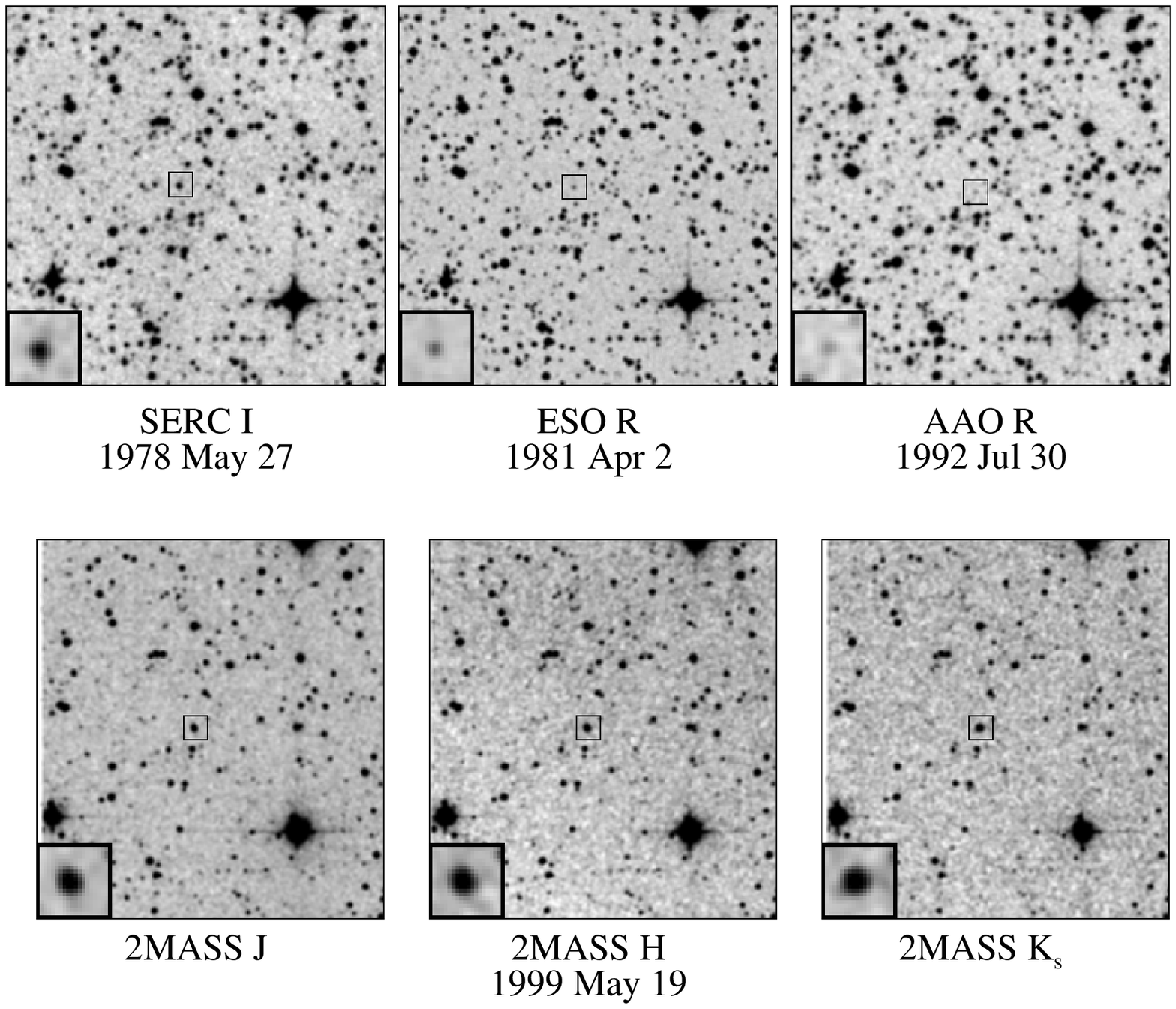}
\caption{Field images of {\name} from SERC $I_N$ (top left), ESO $R$
(top middle) and AAO $R$ (top right) photographic plates; and from
2MASS (bottom).
All images are scaled to the same resolution and oriented with north up and east to the left.  Photographic plate images are 5$\arcmin$ on a side.  
Inset boxes 20$\arcsec$$\times$20$\arcsec$ in size indicate the 
position of the source after correcting for its motion, and are expanded
in the lower left corner of each image.
\label{fig_finder}}
\end{figure}

{\namesh} was identified as a relatively bright ($J = 13.23{\pm}0.3$)
and red ($J-K_s$ = 1.33$\pm$0.04) unresolved source imaged twice
by 2MASS on 1999 May 19 and 2001 February 10 (UT).
The difference in the astrometry of this source
between the two imaging epochs implies a proper motion 
$\mu$ = 0$\farcs$72$\pm$0$\farcs$12 yr$^{-1}$
at a position angle $\theta$ = 235$\degr$.
As shown in Figure~\ref{fig_finder}, counterparts of {\namesh}
can be seen in SERC $I_N$ (epoch 1978 May 27 UT)
and ESO $R$ (epoch 1981 April 2 UT) photographic plates,
offset by 16$\farcs$6 and 15$\farcs$2, respectively,
at the positions expected from the 2MASS proper motion.
A marginal source is also present in 
a 1992 July 30 (UT)
AAO $R$-band photographic plate.  SSS measurements of {\namesh} give $I_N$ = 17.04, $R_{ESO}$ = 19.41 and 
$\mu$ = 0$\farcs$74$\pm$0$\farcs$17 yr$^{-1}$.  

To obtain
a more refined measure of its proper motion, we combined
the SSS 
and 2MASS astrometry
of {\namesh} spanning 22.7 yr, as listed in Table~\ref{tab_astrometry}.  
Linear fits to the 
right ascension and declination over this period yield
$\mu_{\alpha}\cos{\delta}$ = $-$0$\farcs$634$\pm$0$\farcs$013 yr$^{-1}$
and $\mu_{\delta}$ = $-$0$\farcs$367$\pm$0$\farcs$012 yr$^{-1}$,
for a combined motion of $\mu$ = 0$\farcs$73$\pm$0$\farcs$03 yr$^{-1}$ 
at $\theta$ = 239$\fdg$9$\pm$1$\fdg$0.  Note that these values do
not take into consideration parallactic
motion.  The high proper motion of this optically-faint red source 
made it a high priority target for follow-up observations.

\section{Imaging Observations}

\subsection{Data Acquisition and Reduction}

{\namesh} was imaged again in the near-infrared
on 2006 April 8 (UT) using SpeX on IRTF.
Conditions during this night were poor, with patchy cirrus and clouds.
Seeing was 0$\farcs$8--0$\farcs$9 at $J$-band.
Acquisition images with the instrument's guiding camera immediately
revealed
two distinct sources at the position of {\namesh},
separated by roughly 1$\arcsec$.  After rechecking telescope focus,
we obtained a series of 20s exposures of the pair 
in the $J$, $H$ and $K$ filters,\footnote{$JHK$ filters on SpeX
are based on the Mauna Kea Observatory near-infrared (MKO-NIR)
filter set \citep{sim02,tok02}.}
interspersed with images of a nearby,
bright point source 2MASS~J15195960$-$4421523
($J$ = 13.20$\pm$0.03, $J-K_s$ = 0.72$\pm$0.04) as a
point spread function (PSF) calibrator.
Four dithered exposures were obtained for each source and filter,
for a total of 80~s integration each.
We also obtained twilight and bias exposures in all three
filters on 2006 April 11 (UT)
during the same observing campaign for pixel response calibration.

\begin{figure}
\epsscale{1.0}
\plotone{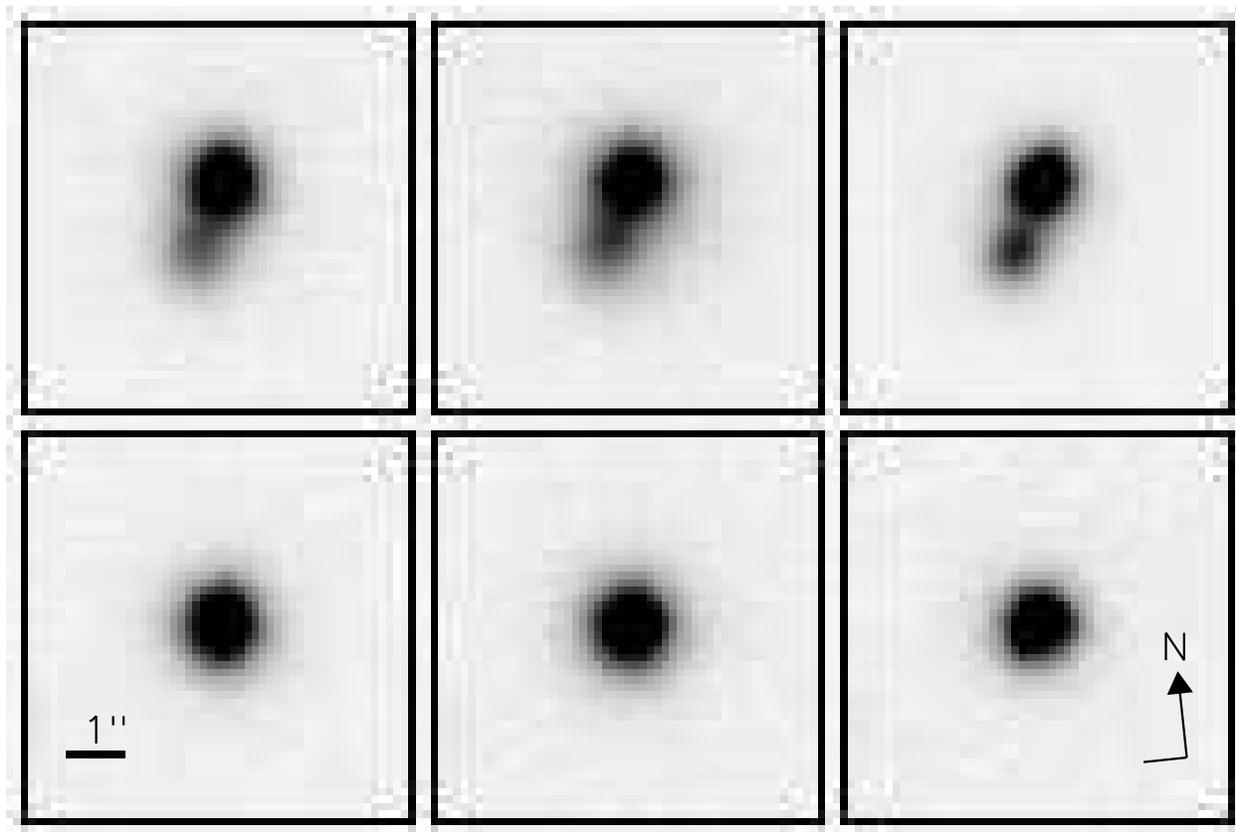}
\caption{SpeX images of {\name}AB (top) and the PSF calibrator
2MASS~J15195960$-$4421523 (bottom) at $J$ (left), $H$ (middle) and $K$ (right).
Images are 6$\arcsec$ on a side.  The image scale is indicated in the bottom left image, and the orientation of all six images (6$\fdg$4$\pm$0$\fdg$3)
is indicated in the bottom right image.
\label{fig_resolve}}
\end{figure}

The imaging data were cleaned using a pixel
mask constructed from the bias and twilight images,
then pair-wise subtracted and divided by a normalized flat field 
frame constructed from the median-combined, bias-subtracted twilight observations (appropriate to the respective filter).
The calibrated images were centered and coadded
to produce a mosaic for each filter/pointing combination.
Figure~\ref{fig_resolve} displays $6{\arcsec}{\times}6{\arcsec}$
subsections of these mosaics for the target and PSF star.
Two sources are clearly resolved at the position of {\namesh}, 
lying along a 
northwest-southeast axis. The northernmost source is 
markedly brighter in all three bands, particularly at $J$ and $H$.

\subsection{PSF Fitting}

Relative astrometry and photometry of the {\namesh} pair were determined
using the same iterative PSF fitting algorithm described in \citet{me06d2200}.
 For each filter, we fit centered
6$\arcsec\times$6$\arcsec$ subsections of each pair-wise subtracted image
of {\namesh} to equivalent
subsections for all four PSF images, a total of 16 fits per filter.
The fitting algorithm first determined an initial estimate of the peak position
and fluxes of the two components, then recursively matched model images (constructed from the PSF frames) to the data.
The primary pixel coordinates, secondary pixel
coordinates, primary flux and secondary flux in the model image were adjusted iteratively
in steps of 0.1 pixels and 0.01 fraction flux
until residuals were minimized, of order 1--2\% of the source peak flux.
The mean of the flux ratios\footnote{Note that the imaging observations
were taken in non-photometric conditions; however, we assume that the
relative atmospheric transmission over the small separation of the {\namesh} pair
was constant, and that any differential 
atmospheric absorption has minimal effect
on the relative flux within each individual filter.}, separations ($\rho$) and position angles ($\phi$) for each filter are 
listed in Table~\ref{tab_properties}.
Uncertainties include scatter in the PSF fit measurements,
and values take into account the pixel scale (0$\farcs$120$\pm$0$\farcs$002 pixel$^{-1}$) and rotator position (6$\fdg$4$\pm$0$\fdg$3; J.\ Rayner, 2005, private communication) of SpeX during the observations.  
As in \citet{me06d2200}, we found no systematic offsets
in our fitting results based on PSF simulations.

To determine component magnitudes, we combined our relative flux measurements
with composite photometry from 2MASS.
Because relative fluxes were measured
using MKO-NIR filters, we first derived
the filter translation factors $\delta_{M} = M^{2MASS}-M^{MKO}$ for each component,
where ${\Delta}M^{2MASS} = {\Delta}M^{MKO} + \delta_{M,B} - \delta_{M,A}$
as described in \citet{mce06}.  These factors were calculated by convolving
filter response curves (filter transmission function 
$\times$ optical response 
$\times$ telluric absorption at an airmass of 1)
for 2MASS \citep[see Cutri et al.\ (2003), $\S$~IV.4.a]{coh03}
and MKO-NIR filters (S.\ Leggett 2004, private communication)
with the individual component spectra ($\S$~4.1) and a spectrum
of the A0~V star Vega \citep{hay85,mnt85}.
These calculations
yield $\delta_{J,B}-\delta_{J,A}$ = 0.018, 
$\delta_{H,B}-\delta_{H,A}$ = $-$0.001 and 
$\delta_{K,B}-\delta_{K,A}$ = $-$0.001.  
The filter translation factors are therefore negligible, likely due to the similarity of the 
component spectra in the regions sampled by the $JHK$ filters, and
were disregarded.  Component magnitudes are 
listed in Table~\ref{tab_component}.

\subsection{Common Proper Motion}

The high proper motion of {\namesh} and the brightness of its components
allow us to immediately deduce common proper motion for this pair.  
Both components are bright enough
to be detected individually by 2MASS.  Since the earliest 
2MASS image of this field was taken 6.89 yr prior to the 
SpeX observations, at least one of these sources would have 
moved a full 5$\farcs$05 from the SpeX position, easily resolvable by 2MASS.
Since no secondary source is seen in these earlier images (Figure~\ref{fig_finder}), implying a limiting
separation of $\sim$1$\farcs$5 \citep{me05gl337cd},
we conclude that the two sources are comoving to within 
$\sim$0$\farcs$2 yr$^{-1}$ of motion and to within 
$\sim$15-20$\degr$ in position angle.  These limits are sufficiently stringent
that common proper motion can be confidently claimed,
and we refer to this pair hereafter as {\namesh}AB.

\section{Spectroscopic Observations}

\subsection{Prism Spectroscopy}

\begin{figure}
\epsscale{0.8}
\plotone{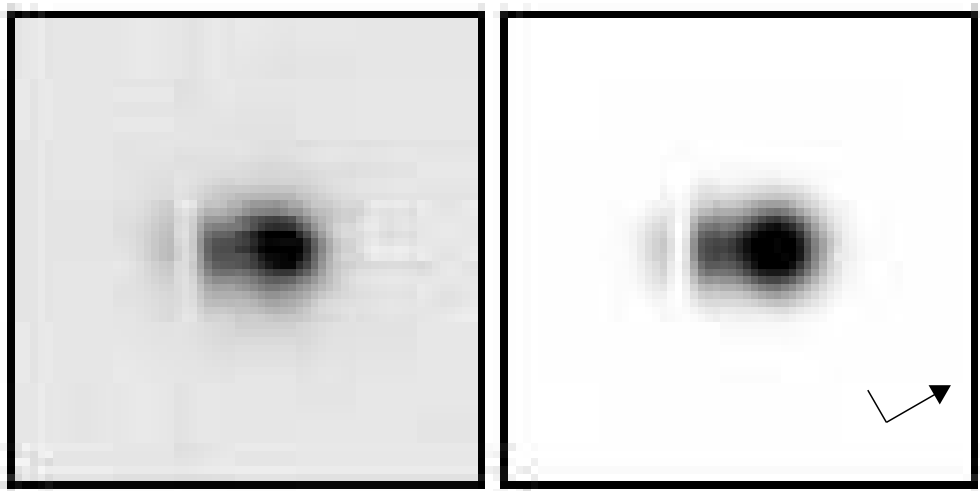}
\caption{{\em Left}: Stacked $K$-band images of {\namesh} during 
acquisition of the fainter component.  The brighter component was
used for guiding, while the secondary was offset slightly from the
slit center to minimize contamination. {\em Right}: Model of the guider
image used
to compute differential refraction (DR) and contamination effects.
Both images are 7$\farcs$2 on a side and orientation on the sky is 
indicated by the arrow.
\label{fig_slitim}}
\end{figure}

Resolved, low resolution near-infrared
spectral data for the {\namesh} components were
obtained on 2006 April 8 (UT), also using SpeX.
We used the 0$\farcs$3 slit to better separate flux from each
component, yielding 0.75--2.5~$\micron$
spectra with resolution {\ldl} $\approx 250$
and dispersion across the chip of 20--30~{\AA}~pixel$^{-1}$.
To obtain separate spectra for each component, we set 
the image rotator to 300$\degr$ so that
one source lay in the slit while the other source was positioned
orthogonal to the slit axis and used for guiding
(Figure~\ref{fig_slitim}).
This orientation was not aligned with the parallactic angle 
($\sim$15$\degr$), and as the observations were made at fairly large airmass
(2.42--2.60),
differential refraction (DR) effects are expected \citep[see below]{fil82}. 
Six and eight exposures of 90~s each were obtained for the A and B components, 
respectively, both in an ABBA dither pattern.
The A0~V star HD~133820 was observed immediately
afterward, at a similar airmass (2.66) and with the slit aligned to the parallactic angle, for flux calibration.
Internal flat field and Ar arc lamps were then observed
for pixel response and wavelength calibration.

Data were reduced using the SpeXtool package, version 3.3
\citep{cus04} using standard settings.
First, the images were corrected for linearity, pair-wise subtracted, and divided by the
corresponding median-combined flat field image.  Spectra were optimally extracted using the
default settings for aperture and background source regions, and wavelength calibration
was determined from arc lamp and sky emission lines.  Multiple spectral observations for each
source were then median-combined after scaling the individual
spectra to match the highest signal-to-noise
observation.  Telluric and instrumental response corrections for the science data were determined
using the method outlined
in \citet{vac03}, with line shape kernels derived from the arc lines. 
Adjustments were made to the telluric spectra to compensate
for differing \ion{H}{1} line strengths in the observed A0~V spectrum
and pseudo-velocity shifts.
Final calibration was made by
multiplying the observed target spectrum by the telluric correction spectrum,
which includes instrumental response correction through the ratio of the observed A0~V spectrum
to a scaled, shifted and deconvolved Kurucz\footnote{\url{http://kurucz.harvard.edu/stars.html}.}
model spectrum of Vega. 

Corrections for DR effects 
in the spectrum of the fainter component, both in terms of 
slit losses and contamination from the bright primary,
were determined by simulating the light throughput for both sources 
as a function of wavelength.
We first created a model of stacked $K$-band guiding images 
(acquired at one dither position during the spectral observations) using 
two symmetric Gaussian surfaces and a box-car slit profile.
The width of the Gaussian profiles
were determined by one-dimensional Gaussian fits of image slices through
the brighter component parallel
to the slit axis.  Note that these values are slightly larger ($\sim$1$\arcsec$)
than the measured seeing due to telescope guiding jitter. 
We adjusted the fluxes and pixel offsets (perpendicular to the slit)
of the two components in our model using an iterative
algorithm similar to that used for the PSF fitting,
minimizing residuals with respect to the $K$-band guider image.
Figure~\ref{fig_slitim} displays the best fit model; note that the
the secondary was actually centered slightly off the slit.
We then calculated DR pixel shifts as a function of wavelength 
using the formalism of
\citet[see also Schubert \& Walterscheid (2000)]{roe02}, assuming standard
atmospheric pressure and temperature, a relative humidity of 50\%
and a true zenith distance of 55$\fdg$5.  The calculated
shifts move both components
closer to the slit center at wavelengths shorter than 2.2~$\micron$, by as much as 1.5 pixels at 1.2 $\micron$.  Hence,
slit losses are greatest at the longest wavelengths
while contamination from the primary is greatest at the shortest
wavelengths.
Both effects were calculated using our model images by 
shifting the centers of the Gaussian profiles in accordance with the DR shifts 
(keeping the seeing and relative flux ratios constant), and calculating the
total light throughput for both components in the slit.  
We found slit losses to be roughly 20\% at $K$ relative to $J$, 
while light contamination 
ranged from 3\% at $K$ to $>$10\% for $\lambda < 1.2 \micron$.

\begin{figure}
\epsscale{0.8}
\plotone{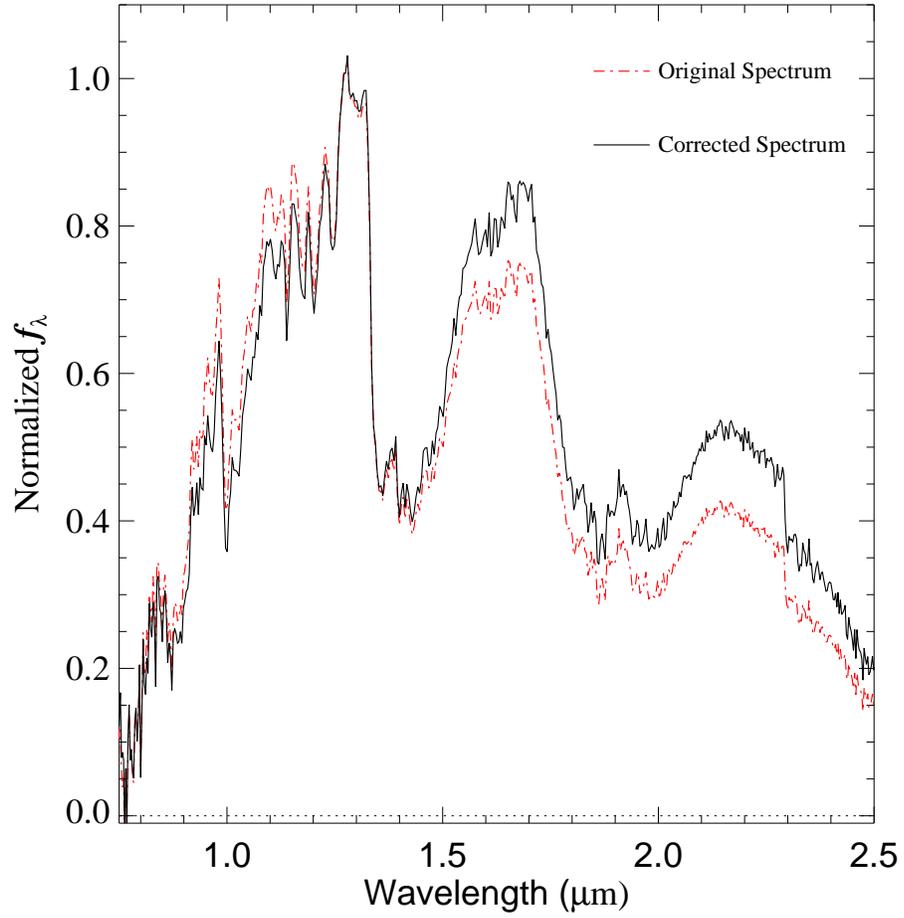}
\caption{SpeX prism spectrum of {\namesh}B before (red dash-dot line) 
and after (black solid line) correction of differential
refraction (DR) and contamination effects.  Both spectra are normalized
at 1.28~$\micron$.  \label{fig_corrected}}
\end{figure}

The spectrum of {\namesh}B was corrected for slit losses using that
component's model-derived relative throughput values.  Contamination from the primary
was removed by subtracting a spectrum of {\namesh}A scaled by the
wavelength-dependent contribution of its light through the slit relative
to the secondary from the DR model. 
Figure~\ref{fig_corrected} shows the spectrum of {\namesh}B
before and after these effects have been accounted for.  
The corrected spectrum
is significantly redder, as both slit losses and light contamination
suppress $K$-band light from this component.  
Similar simulations were used to 
calculate DR slit losses for the spectrum of the
primary assuming it to be centered on the slit at $K$-band. 
These were smaller ($<$ 10\% slit loss for $\lambda >$ 1 $\micron$),
due to the large size of the seeing disk relative to the 
slit width. Light contamination from the secondary was found to be
negligible.

\begin{figure}
\epsscale{0.8}
\plotone{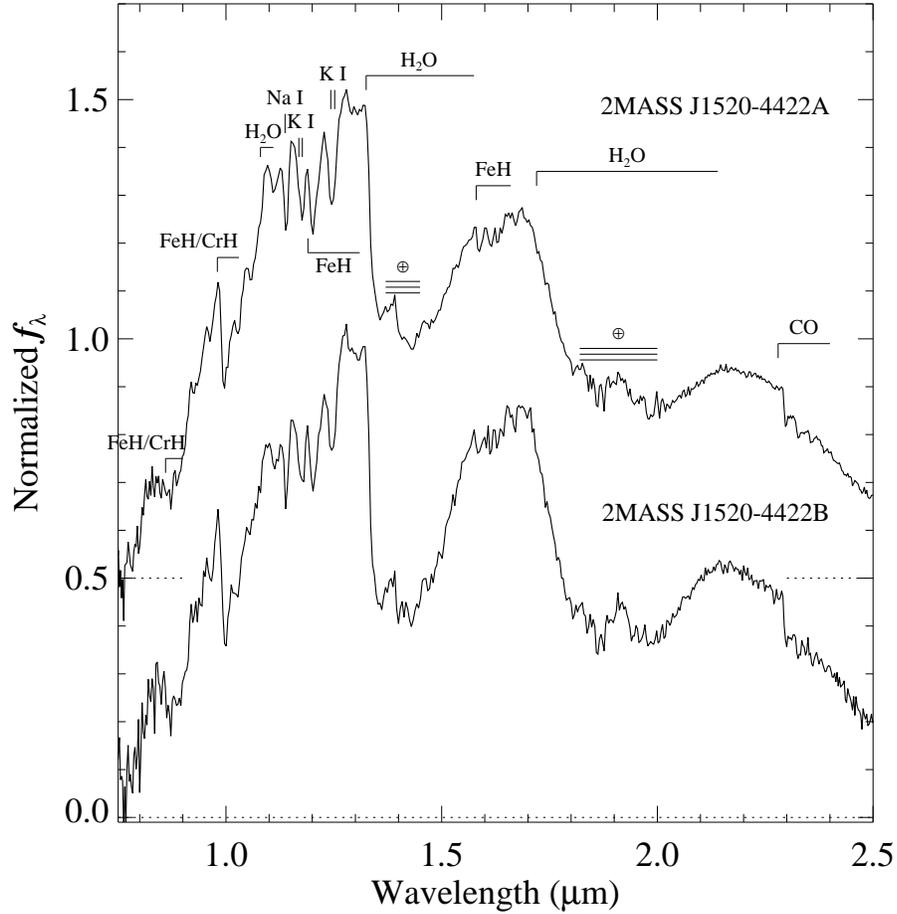}
\caption{Near-infrared SpeX spectra of {\namesh}A (top) and B (bottom).
Data are normalized at 1.28~$\micron$, with {\namesh}A offset
by a constant for clarity (dotted lines).  Major molecular (FeH, CrH, {\water}, CO) and atomic
(\ion{Na}{1} and \ion{K}{1}) absorption features are labelled,
as well as regions of strong telluric absorption ($\oplus$).
\label{fig_prism}}
\end{figure}

The reduced prism spectra (including DR and contamination corrections)
of the two {\namesh} components
are shown in Figure~\ref{fig_prism}.  
These spectra are qualitatively similar to JH spectra presented 
in \citet{ken06}.
Both components have spectral energy
distributions consistent with early-type L dwarfs,
including red optical and near-infrared spectral slopes;
deep {\water} absorption bands at 1.4 and 1.9~$\micron$;
strong CO absorption at 2.3~$\micron$; FeH and CrH bands
at 0.86, 0.99, 1.2 and 1.6~$\micron$;
a notable absence of TiO and VO features shortward of 1 $\micron$
and unresolved \ion{Na}{1} and \ion{K}{1} atomic lines at $J$-band
\citep{rei01a,tes01,mcl03,nak04}.
The 0.99 $\micron$ FeH band in {\namesh}B is unusually strong, although
this may be due to uncorrected DR effects.
The stronger {\water} and FeH molecular bands of {\namesh}B,
and its redder near-infrared color, indicate that it is 
of later type than {\namesh}A.

\subsection{Cross-dispersed Spectroscopy}

Higher resolution cross-dispersed (XD) spectra for the composite system
(i.e., both sources simultaneously in the slit) were obtained on 2006 April 11 (UT).  Conditions on this night were somewhat improved,
with light cirrus and moderate seeing (0$\farcs$8 at $J$-band). 
The 0$\farcs$5 slit was used for
a spectral resolution {\ldl} $\approx 1200$ and dispersion across the chip of
2.7--5.3~{\AA}~pixel$^{-1}$.  The slit was aligned to the parallactic angle
and 8 exposures (4 of 250 s, 4 of 300 s) dithered
in an ABBA pattern were obtained for a total integration time of 2200 s.  HD~133820 was again observed immediately after {\namesh} and at a similar airmass (2.75),
followed by calibration lamps.  As with the prism data, XD data were
reduced using the SpeXtool package, following similar
procedures but using a line shape kernel
derived from the 1.005 $\micron$ \ion{H}{1} Pa $\delta$ 
line in the A0~V calibrator spectra. 
After telluric and flux calibration,
the five orders spanning 0.82--2.43 $\micron$ 
were scaled and combined using the prism spectrum of {\namesh}A
as a relative flux template.
 
\begin{figure}
\epsscale{0.9}
\plotone{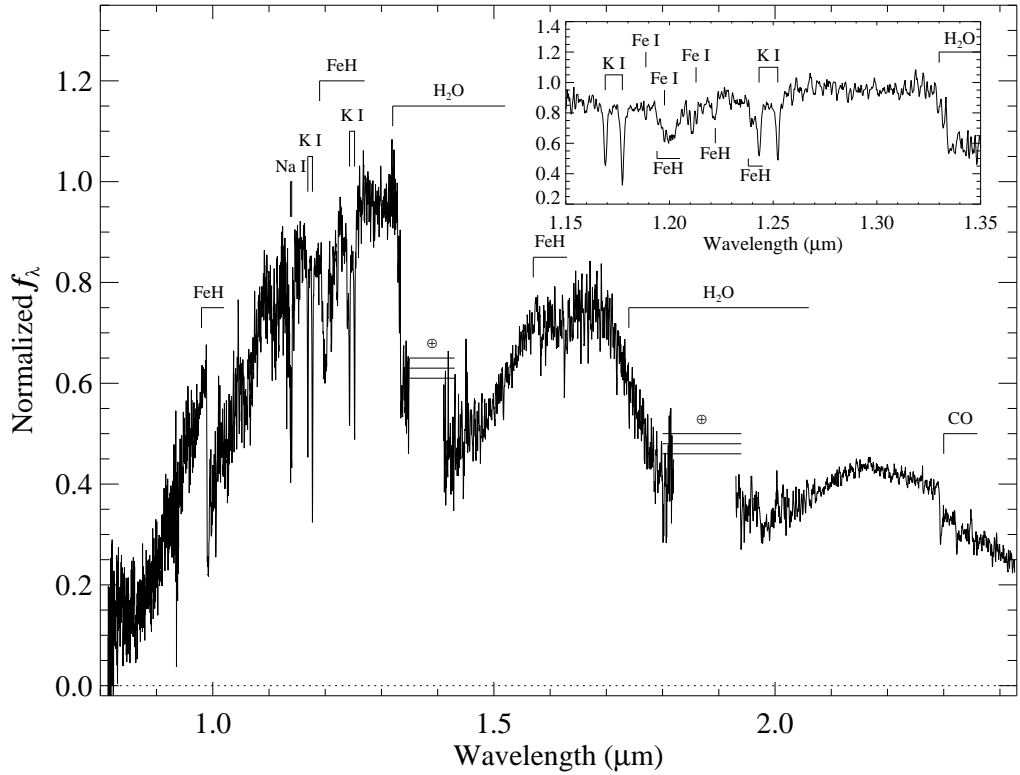}
\caption{Cross-dispersed spectrum of the composite {\namesh} system.
Data are normalized at 1.28~$\micron$.  Major molecular and atomic
absorption features are indicated,
as well as regions of strong telluric absorption ($\oplus$).
The inset shows a close-up view of the 1.15-1.35 $\micron$ region,
hosting several \ion{K}{1} and \ion{Fe}{1} lines and FeH
and {\water} molecular bands.
\label{fig_sxd}}
\end{figure}

The reduced XD spectrum for {\namesh} is shown in Figure~\ref{fig_sxd}.
While these data have lower
overall signal-to-noise than the prism data (particularly shortward
of 0.98 $\micron$ and in the telluric bands),
many of the small wiggles observed in this spectrum are real, 
arising largely from
{\water} and FeH transitions and numerous atomic metal lines.
The strongest atomic features are the \ion{Na}{1} and \ion{K}{1} doublet
lines at $J$-band, which are shown in detail in the inset of Figure~\ref{fig_sxd}.  In addition, there are several lines from \ion{Fe}{1} and
FeH, and broader FeH bands, in this spectral region.  
CO bandheads at 2.299, 2.328, 2.357 and 
2.388 $\micron$ are clearly discerned, but the 2.206/2.209 
$\micron$ \ion{Na}{1} doublet is notably absent.
The absence of this feature provides further evidence that {\namesh} 
is composed of L dwarfs \citep{mcl03,cus05}.  Similarly, 
the absence of the 1.314 $\micron$ \ion{Al}{1} line
is consistent with an L spectral
type, while the presence of the 1.189 $\micron$ \ion{Fe}{1} line suggests
a composite spectral type of L3 or earlier \citep{mcl03}.

Equivalent widths (EW) of the 1.169, 1.178 and 1.253 $\micron$ \ion{K}{1} lines
and 1.189 \ion{Fe}{1} line were measured using 
Gaussian fits to the line profiles and linear fits to the nearby
pseudocontinuum (the 1.244 $\micron$ \ion{K}{1} was not measured as its blue
wing is blended
with an FeH band).  Values of 6$\pm$2, 8.3$\pm$1.8, 8.0$\pm$1.6
and 0.5$\pm$0.2 {\AA}, respectively, were derived and
are comparable to other early-type L 
dwarfs \citep{cus05}.

\section{Analysis}

\subsection{Spectral Classification and Estimated Distance}

The resolved prism spectra of the {\namesh} components allow
us to derive individual spectral types.  However,
while a robust classification scheme for L dwarfs
exists at optical wavelengths \citep{kir99,kir00},
there is as yet no well-defined scheme 
in the near-infrared; i.e., following the tenets of the MK Process
\citep{mor43,cor94}.
We therefore determined subtypes for these sources 
using a variety of spectral
indices from the literature that are applicable to low
resolution near-infrared spectral data.
\citet{rei01a} have defined several indices measuring the strengths
of {\water} and CO bands, as well as color ratios, for late-type M and L dwarf spectra.
They provide linear spectral type
calibrations anchored to optical classifications
for their {\water}-A and {\water}-B indices
(sampling the 1.1 and 1.4~$\micron$ {\water} bands) and the $K1$ index
of \citet{tok99} over spectral types M8 to L6.
\citet{tes01} defined six indices sampling {\water}
bands and color ratios for spectral data of comparable
resolution to our SpeX prism observations, and provide linear 
calibrations for these indices over types L0 to L8.   
\citet{me02a} also defined {\water} band indices
and color ratios, as well as {\meth} band indices, in their 
near-infrared classification scheme for T dwarfs; three of the {\water}
indices show good correlation with M5--L7 optical spectral type,
and one {\meth} index correlates with L3--T3 spectral type.
Finally, \citet{geb02} utilized several near-infrared indices to classify late-type M, L and T dwarfs, two of which  
({\water} 1.5$\micron$ and {\meth} 2.2$\micron$) are applicable
in the L dwarf regime (for spectral types L0--L9 and L3--L9, respectively).  
Unlike the other near-infrared 
schemes, \citet{geb02} tie spectral types to their indices with 
predetermined numerical ranges
set by measurements for a large sample of VLM dwarf spectra.

Table~\ref{tab_classify} lists the values and associated spectral
types of these indices for each component spectrum prism 
of {\namesh}, as well as the 
composite XD spectrum.  Note that we 
have not included color indices from \citet{tes01} and \citet{me02a} due
to possible residual DR effects.  There is fairly consistent agreement
in the derived spectral types between these schemes, likely due to the 
similarity in features used to define the indices (primarily {\water} bands).
Mean subtypes for all indices
are L1.5 and L4.5 for {\namesh}A and B, 
with scatter of 0.5 and 0.8 subtypes, respectively.  This is consistent
with the types derived by \citet{ken06}, 
L2 and L4, within the stated uncertainties. The XD 
spectrum is classified L2.5$\pm$0.8, as expected for contributions from
both components to the combined light spectrum.

\begin{figure}
\epsscale{1.0}
\plotone{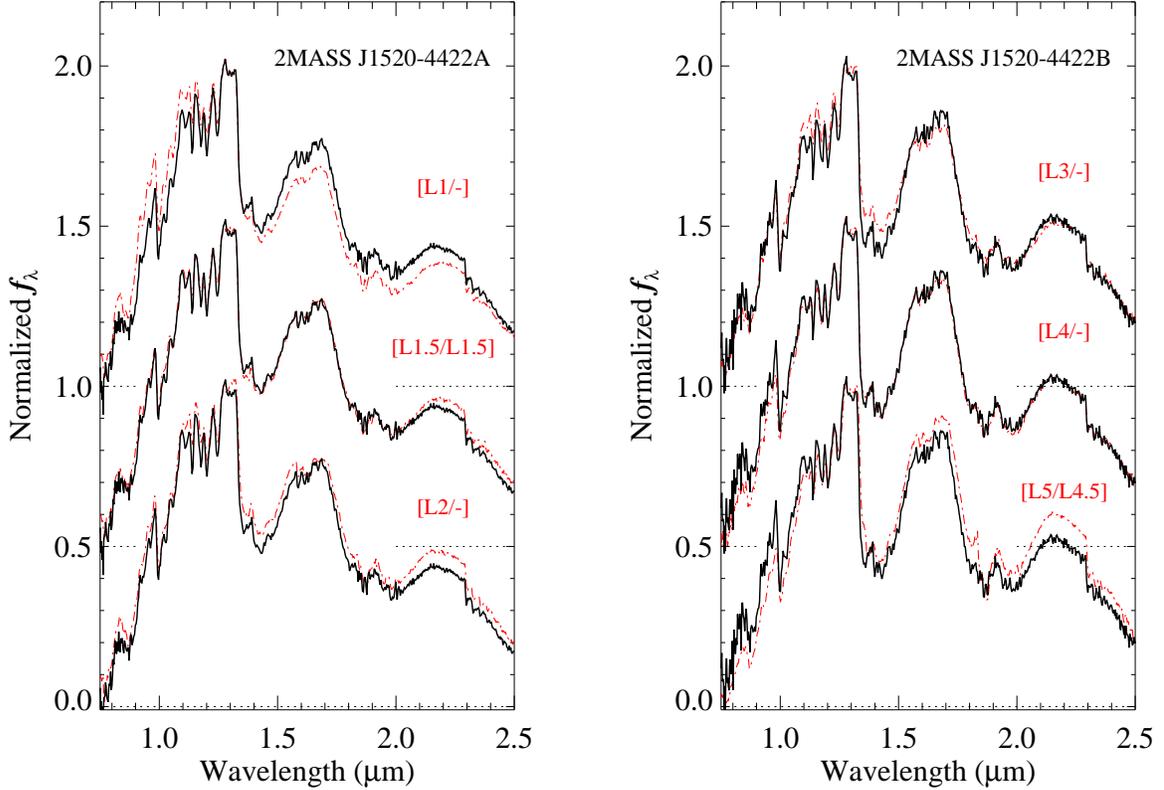}
\caption{Comparison of the near-infrared spectra of {\namesh}A and B
(black solid lines in left and right
panels, respectively)
to equivalent data (red dash-dot lines) for the spectral comparison stars.
In the left panel we compare {\namesh}A to  
2MASS~J14392836+1929149 (L1 optical classification), 2MASS J20575409-0252302 (L1.5 optical and near-infrared classification)
and SSSPM J0829-1309 (L2 optical classification).   In the right panel
we compare {\namesh}B to SDSS J202820.32+005226.5 (L3 optical classification),
2MASS J11040127+1959217 (L4 optical classification) and
GJ 1001BC (L5 optical classification, L4.5 near-infrared classification).
All spectra have been normalized
at 1.28~$\micron$ and offset by constants (dotted lines).
\label{fig_classify}}
\end{figure}

To confirm these classifications, we compared the prism spectra
to an array of equivalent data for previously classified field
L dwarfs with similar optical and/or near-infrared subtypes, as shown in
Figure~\ref{fig_classify}.  The spectrum of
{\namesh}A is nearly identical to that of 2MASS J20575409-0252302, classified
L1.5 from both optical \citep{cru03} and near-infrared \citep{ken04} data;
and appears to be intermediate between those of optically-classified L1 
(2MASS J14392836+1929149; \citet{kir99} spectral standard)
and L2 (SSSPM J0829-1309; \citet{sch02}) comparison sources.
Similarly, the spectrum of {\namesh}B shows remarkable agreement with that of the optically-classified L4 2MASS J11040127+1959217 \citep{cru03},
and less agreement with those of optically-classified 
L3 (SDSS J202820.32+005226.5; \citet{haw02}) and 
L5 dwarfs (GJ 1001BC; \citet{gol99}).  These comparisons would appear to verify the index-based classifications from Table~\ref{tab_classify}, and indicate
that the DR and light contamination corrections applied 
to the spectrum of {\namesh}B were accurate.

\begin{figure}
\epsscale{0.8}
\plotone{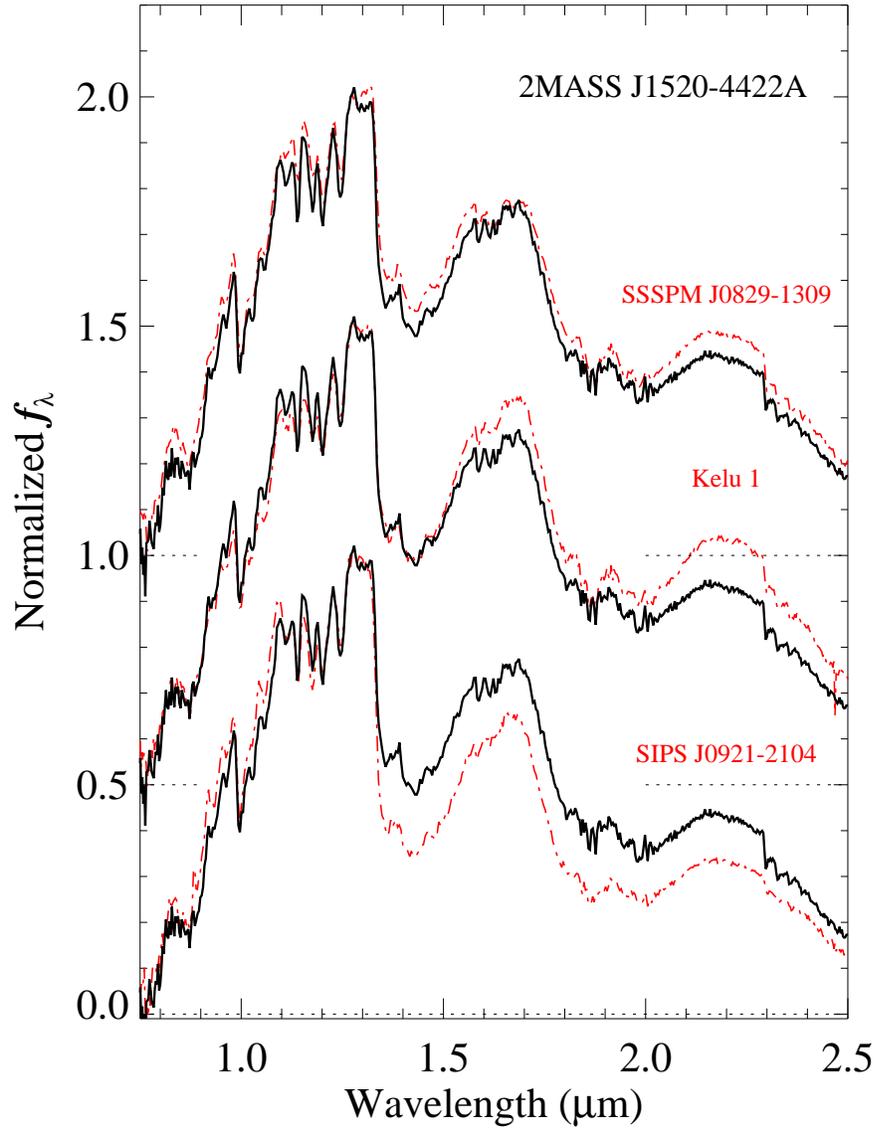}
\caption{Comparison of the near-infrared spectra of {\namesh}A
(black solid lines)
to equivalent data (red dash-dot lines) for three optically-classified L2 dwarfs:
SSSPM J0829-1309, Kelu~1 and SIPS J0921-2104.
All spectra have been normalized
at 1.28~$\micron$ and offset by constants (dotted lines).
\label{fig_L2comp}}
\end{figure}

The agreement of the near-infrared spectra of {\namesh}A and B with those of 
optically-classified sources suggests that the derived
subtypes could be adopted as their optical classifications.
However, a number of studies \citep{ste03,kna04,kir05} have found
that optical and near-infrared
spectral types cannot be assumed to be identical for mid- and late-type 
L dwarfs due to the complexity of their atmospheres.
Figure~\ref{fig_L2comp} illustrates that this is an issue even amongst
early-type L dwarfs, comparing SpeX prism data
for three optically-classified
L2 dwarfs --- SSSPM J0829-1309, Kelu~1 \citep{rui97} and SIPS J0921-2104 
\citep{dea05} --- to that of {\namesh}A.  Despite their equivalent
optical classifications, these three sources have very different
near-infrared spectra.  Both SSSPM J0829-1309 and Kelu~1 appear to
have redder 1-2.5 $\micron$ spectra
than {\namesh}A, while SIPS J0921-2104 is bluer.  SSSPM J0829-1309 has
weaker 1.4 and 1.7 $\micron$ {\water} absorption while these same bands are
markedly deeper in the spectrum of SIPS J0921-2104.  
The spectral indices used above
yield mean near-infrared subtypes of L1, L2.5 and L4 for 
SSSPM J0829-1309, Kelu~1 and SIPS J0921-2104, respectively
(see also Knapp et al.\ 2004 and Lodieu et al.\ 2005 regarding
the near-infrared spectrum of Kelu~1).
Such classification disagreements have been attributed to a variety
of causes, including variations in photospheric dust content
\citep{ste01,ste03}, unresolved multiplicity \citep{me0423,liu05}
and gravity effects \citep{mcg04,kir06}.
These complexities emphasize the need for a robust, independent
and multi-dimensional
near-infrared classification scheme for L dwarfs.

Assuming that the derived near-infrared 
spectral types for {\namesh}A and B are at least
suitable proxies for their optical
classifications, spectrophotometric distance estimates for each component 
were computed using
measured photometry and absolute 2MASS magnitude/spectral type relations from
\citet{dah02,cru03,tin03}; and \citet{vrb04}.
These yield mean distances of 19.1$\pm$0.8
and 19.1$\pm$0.7 pc for {\namesh}A and B, respectively, assuming
no uncertainty in the spectral types.  The agreement in the
estimated distances for both components and their common proper
motion are consistent with these two sources being gravitationally bound.
We adopt a mean distance of 
d$_{est}$ = 19$\pm$2 pc, which includes a 0.5 subtype uncertainty in the 
classifications.

\subsection{Radial Velocity and Kinematics}

The large proper motion of {\namesh} and its estimated distance
imply a large tangential motion, 
$V_{tan} = 4.74{\mu}$d$_{est}$ = 66$\pm$7~{\kms}.
This is on the high end of the L dwarf $V_{tan}$ distribution
of \citet{vrb04}, in which only 3/33 = 9$^{+8}_{-3}$\% of the sources
examined have $V_{tan} > 60$~{\kms}.  Similarly, 
only 13$^{+6}_{-3}$\% of late-type M and L dwarfs
in the sample of \citet{giz00} have $V_{tan} > 60$~{\kms}.
These studies suggest that {\namesh} may be an unusually high
velocity system.

To determine its three-dimensional space motion, we
used the XD spectrum to measure a systemic radial
velocity ($V_{rad}$).  Despite the coarse velocity resolution of these data
(${\Delta}V \approx 250$~{\kms}), a relatively accurate determination
of $V_{rad}$ can be obtained
through cross correlation techniques.
We compared the spectrum of {\namesh} to equivalent SpeX XD data 
from \citet{cus05} for four
early/mid-type L dwarfs with measured radial velocities 
\citep{bai04}: 
2MASS J11463449+2230527 (L3; 21$\pm$2~{\kms}),
2MASS J14392836+1929149 (L1; -26.3$\pm$0.5~{\kms}),
2MASS J15074769-1627386 (L5; -39.3$\pm$1.5~{\kms}) and
2MASS J22244381-0158521 (L4.5; -37$\pm$3~{\kms}).
The wavelength scale for the {\namesh} data was shifted
by velocities ranging over -300 to 300~{\kms}
in steps of 1~{\kms}. Then, for each shifted spectrum, the 
cross correlation\footnote{We computed 
$C(V) \propto \int_{\lambda_1}^{\lambda_2}(f^T_{\lambda}(V)-\bar{f^T_{\lambda}})(f^C_{\lambda}-\bar{f^C_{\lambda}})d{\lambda}$, where $f^T_{\lambda}(V)$ is the spectrum of
{\namesh} shifted to velocity $V$, $f^C_{\lambda}$ is the spectrum of the
comparison source, and $\bar{f_{\lambda}}$ is the mean flux over the spectral
band spanning $\lambda_1$ to $\lambda_2$.  The radial velocity of {\namesh}
was determined as $V_{rad} = V_{max}-V_C$, where $V_{max}$ is the velocity that maximizes $C(V)$ and $V_C$ is the known radial velocity of the comparison source.} was computed against each of the comparison spectra in
three wavebands with strong features: the 1.16-1.26 $\micron$
region, hosting numerous \ion{K}{1}, \ion{Fe}{1} and FeH absorptions 
(see inset of Figure~\ref{fig_sxd}); the 1.32--1.34 $\micron$ region, spanning the 1.33 $\micron$ {\water} bandhead; and the 2.290-2.295 $\micron$ region, spanning the 2.292 {\micron} CO bandhead.  After correcting for the radial
velocities of the comparison sources, we derived a mean
$V_{rad} = -60{\pm}12$~{\kms}
for {\namesh}, where the uncertainty is the scatter of 
the 12 cross correlation measurements.  As a secondary check, we 
measured the central wavelengths of the well-resolved 1.169, 1.177, 1.244
and 1.253 $\micron$
\ion{K}{1} lines using Gaussian fits to the line cores and
vacuum wavelengths as listed in the Kurucz Atomic Line Database\footnote{Obtained
through the online database search form created by C.\ Heise and maintained
by P.\ Smith; see 
\url{http://cfa-www.harvard.edu/amdata/ampdata/kurucz23/sekur.html}.} 
\citep{kur95}.  These four lines yield a mean $V_{rad} = -79{\pm}24$~{\kms},
where the uncertainty includes systematic errors as determined from 
similar measurements for the four
L dwarf comparison spectra used above.  
These two measures are consistent, and we adopt a weighted average
of $V_{rad} = -70{\pm}18$~{\kms}.
Note that corrections for Earth's motion 
in the Solar frame of reference have not been included in this value, 
as these corrections are significantly smaller than
our estimated uncertainty.

The total space velocity of {\namesh} is quite high, 
nearly 100~{\kms} relative to the Sun.  Combining position,
proper motion and radial velocity measurements with 
our estimates for the distance of {\namesh} yields local standard of 
rest (LSR) velocity components of $(U,V,W) \approx$ (50,30,-20)~{\kms},
where we have assumed $(U,V,W)_{\sun}$ = (10,5,7)~{\kms} \citep{deh98}.
These values lie just outside of the 1$\sigma$
velocity dispersion sphere of local disk M dwarfs
([$\sigma_U$,$\sigma_V$,$\sigma_W$] $\approx$ [40,28,19]~{\kms}
centered at [$-13,-23,-7$]~{\kms}; Hawley, Gizis \& Reid 1996), 
suggesting old disk kinematics but not necessarily 
membership in the thick disk or halo populations.
The absence of any distinct low metallicity features in the 
near-infrared spectra of the {\namesh} components that characterize
L subdwarfs \citep{me1626} also indicates that
this system is likely an old member of the Galactic disk population.

\subsection{Additional Properties of the {\namesh} Binary}

As early/mid-type L dwarfs, the components of {\namesh} are likely to have
masses close to or below the hydrogen burning minimum mass,
M $\approx$ 0.075~M$_{\sun}$ \citep{cha00,bur01}.
However, because brown dwarfs cool over time, their
{\teff}s and spectral types are a function of both mass and age,
which are difficult to disentangle for individual field objects.
In Table~\ref{tab_component}, we list estimated masses for 
{\namesh}A and B for ages of 1, 5 and 10~Gyr,
assuming {\teff}s of 2200 and 1740~K, respectively
(appropriate for L1.5 and L4.5 dwarfs; \citet{vrb04})
and the evolutionary models of \citet{bur97}.
The kinematics of this system suggest an age older than 1~Gyr,
implying masses of $\gtrsim$0.075 and $\gtrsim$0.064 M$_{\sun}$
for the two components, respectively.
Hence, it is likely that {\namesh}A is a low mass star, while
{\namesh}B is either a low mass star or a 
massive brown dwarf.  

Our spectrophotometric distance estimate of {\namesh} 
implies a projected separation of $\rho$ = 22$\pm$2~AU, which is 
relatively wide for a VLM binary.  Indeed, \citet{me06ppv} have found
that 93\% of known VLM binaries have $\rho <$ 20~AU.
However, ``wide'' VLM binaries such as
{\namesh} do not violate the maximum separation/total system mass trend 
of \citet{me03b}, $\rho_{max} = 1400$M$_{tot}^2$~AU,
which appears to be appropriate for nearly all known 
binaries with M$_{tot}$ $\lesssim$ 0.3 M$_{\sun}$.
These are to be distinguished from a few ``ultrawide'' (separations $>$200~AU)
and/or weakly-bound ($V_{esc}$ $<$ 3.8~{\kms}; \citealt{clo03})
VLM binaries now known \citep{cha04,cha05,luh04,bil05,clo06,jay06}, whose
origin and stability remain controversial \citep{mug05}
and a challenge for VLM dwarf formation models \citep{rpt01,bat03}.
Nevertheless, {\namesh} is sufficiently wide that its estimated orbital
period ($\sim$400~yr) is far too long to be useful
for dynamical mass measurements in the near future.

\section{Discussion}

{\namesh} joins a growing list of VLM binaries that are sufficiently
well-separated to allow resolved photometric and spectroscopic studies
from moderate-sized ground-based telescopes (such as IRTF)
without the need for
adaptive optics corrections.  Table~\ref{tab_widebinary} lists the
16 known VLM binaries with angular separations greater than 0$\farcs$5
that fall into this category;
all are resolvable from the ground at optical and near-infrared
wavelengths during conditions of excellent seeing (indeed, most have
been discovered and/or observed entirely with ground-based facilities).
These systems are also among the most interesting VLM sources currently
under investigation.
Epsilon Indi Bab \citep{sch03,mcc04} is composed of the two nearest 
brown dwarfs to the Sun currently known, followed closely by the secondary
of SCR 1845-6537AB \citep{ham04,bil06}.  
The secondary of the 2MASS J12073347-3932540AB system, a member of the
nearby $\sim$8 Myr TW Hydrae association \citep{giz02}, has an estimated
mass in the planetary-mass regime 
($\sim$5 Jupiter masses; Chauvin et al.\ 2004), and both components may 
harbor circumstellar accretion disks \citep{moh03,giz05a,moh06}.
Oph 11AB (2MASS~J16222521-2405139; \citealt{all06}) is an even younger
and wider ($\sim$240~AU) VLM binary whose components 
have masses nearly in the planetary-mass regime \citep{clo06,jay06,all07}.
2MASS J23310161-0406193AB, Epsilon Indi Bab and Gliese 337CD
are all members of higher order multiples, widely separated from 
more massive primaries \citep{giz00,sch03,me05gl337cd}.
DENIS J020529.0-115925AB may itself be a triplet of brown dwarfs \citep{bou05}.

While resolved near-infrared spectroscopy exists for over half of the
systems listed in Table~\ref{tab_widebinary}, 
only three\footnote{\citet{mar06} report resolved optical
spectroscopy for 9 VLM binaries based on {\em Hubble Space Telescope}
observations, 8 of which have separations less than 0$\farcs$5.
Note that these data had insufficient resolution and signal-to-noise
to measure \ion{Li}{1}
EWs $\leq$ 5~{\AA}.} (DENIS J020529.0-115925AB, \citealt{mar06};
2MASS J11011926-7732383AB, \citealt{luh04}; and Oph 11AB, \citealt{jay06}.)
have resolved optical spectroscopy reported to date.  
Such observations are of particular importance for 
M dwarf/L dwarf binaries due to the presence of 6563 {\AA} H$\alpha$ line,
as discussed above, and the 6708 {\AA} \ion{Li}{1} line.   The latter
is a powerful indicator of mass and age for M dwarf/L dwarf field binaries,
as recently pointed out by \citet{liu05}. This species is depleted 
in the atmospheres of
brown dwarfs and low mass stars more massive than $\sim$0.065~M$_{\sun}$
due to fusion reactions in their cores \citep{reb92}.
Hence the detection of this line in the spectrum of a brown dwarf
sets a upper limit for its mass, 
and a corresponding upper limit on its age for a given {\teff}
and assumed evolutionary model.  
For a binary system composed of two coeval brown dwarfs, \ion{Li}{1}
absorption may be present in one, both or neither component, and any of
these three cases can set different constraints on the age of the system.
In the case of {\namesh}AB,
if the 6708~{\AA} \ion{Li}{1} line
is present in the spectra of both components, 
then the evolutionary models of \citet{bur97} predict a system age
of $\lesssim$~600~Myr.  
If \ion{Li}{1} in absent in both component spectra, than the system
is $\gtrsim$~1~Gyr old, consistent with its kinematics.  
The most interesting case is if \ion{Li}{1}
is present only in the spectrum of the secondary, as this would provide
a tight constraint on the age of the system ($\sim$0.6--1~Gyr).
While this ``binary Li test'' has been indirectly used
for sources with composite optical spectroscopy \citep{me0423,liu05,mce06},
the wide separation of {\namesh} makes it an excellent
target for the first unambiguous application of this technique.

\section{Summary}

We have identified a well-resolved L1.5+L4.5 binary, {\namesh}.
This source was found by us as a high proper motion source 
($\mu = 0{\farcs}73{\pm}0{\farcs}03$ yr$^{-1}$)
in the 2MASS point source
catalog, and concurrently by \citet{ken06} 
in the SuperCOSMOS Sky Survey catalog.  
The apparent separation ($\rho = 1{\farcs}174{\pm}0{\farcs}016$)
and estimated distance of this system (19$\pm$2 pc) imply a wide
projected separation (22$\pm$2 AU) that is nevertheless consistent with the maximum
separation/total mass relation of \citet{me03b}.  The {\namesh} system exhibits
a large space velocity relative to the Sun
($V_{tan} =66{\pm}7$~{\kms}, $V_{rad} = -70{\pm}18$~{\kms}), and its
kinematics are consistent with an old Galactic disk dwarf system.
As a source at the boundary of the hydrogen burning mass limit, and with components spanning
the early/mid-type L dwarf regime, {\namesh} is a prime target for resolved
photometric and spectroscopic investigations of magnetic activity,
dust evolution and variability, and rotational velocity trends; and 
a test case for the binary Li test.

\acknowledgments

The authors would like to thank telescope operator Eric Volquardsen,
instrument specialist John Rayner, and Alan Tokunaga for their 
assistance and hospitality during the
observations.  AJB acknowledges useful discussions
with Subhanjoy Mohanty
during the preparation of the manuscript, and helpful comments
from our referee Herve Bouy. 
This publication makes
use of data from the Two Micron All Sky Survey, which is a joint
project of the University of Massachusetts and the Infrared
Processing and Analysis Center, and funded by the National
Aeronautics and Space Administration and the National Science
Foundation. 2MASS data were obtained from the NASA/IPAC Infrared
Science Archive, which is operated by the Jet Propulsion
Laboratory, California Institute of Technology, under contract
with the National Aeronautics and Space Administration.
This program has benefitted from the M, L, and T dwarf compendium housed at DwarfArchives.org and maintained by Chris Gelino, Davy Kirkpatrick, and Adam Burgasser;
and the VLM Binary Archive maintained by N. Siegler at 
\url{http://paperclip.as.arizona.edu/$\sim$nsiegler/VLM\_binaries}.
The authors wish to recognize and acknowledge the 
very significant cultural role and reverence that 
the summit of Mauna Kea has always had within the 
indigenous Hawaiian community.  We are most fortunate 
to have the opportunity to conduct observations from this mountain.

Facilities: \facility{IRTF(SpeX)}

\clearpage

\clearpage

\clearpage

\clearpage

\clearpage

\clearpage

\clearpage

\clearpage

\begin{deluxetable}{lccl}
\tabletypesize{\small}
\tablecaption{Astrometry for {\name}AB. \label{tab_astrometry}}
\tablewidth{0pt}
\tablehead{
\colhead{Epoch (UT)} &
\colhead{$\alpha$\tablenotemark{a}} &
\colhead{$\delta$\tablenotemark{a}} &
\colhead{Source} \\
}
\startdata
1978 May 27 & 15$^h$20$^m$03$\fs$46 & $-$44$\degr$22$\arcmin$34$\farcs$1 & SERC $I_N$ (SSS) \\
1981 Apr 2 & 15$^h$20$^m$03$\fs$34 & $-$44$\degr$22$\arcmin$35$\farcs$6 & ESO $R$ (SSS) \\
1999 May 19 & 15$^h$20$^m$02$\fs$24 & $-$44$\degr$22$\arcmin$41$\farcs$9 & 2MASS \\
2001 Feb 10 & 15$^h$20$^m$02$\fs$14 & $-$44$\degr$22$\arcmin$42$\farcs$7 & 2MASS \\
\enddata
\tablenotetext{a}{Equinox J2000 coordinates.}
\end{deluxetable}

\begin{deluxetable}{lll}
\tabletypesize{\footnotesize}
\tablecaption{Properties of the {\name}AB System. \label{tab_properties}}
\tablewidth{0pt}
\tablehead{
\colhead{Parameter} &
\colhead{Value} &
\colhead{Ref.} \\
}
\startdata
$\alpha$\tablenotemark{a} & 15$^h$20$^m$02$\fs$24 & 1 \\
$\delta$\tablenotemark{a} & $-$44$\degr$22$\arcmin$41$\farcs$9 & 1 \\
$\mu$ & 0$\farcs$73$\pm$0$\farcs$03 yr$^{-1}$ & 1,2,3 \\
$\theta$ & 239$\fdg$9$\pm$1$\fdg$0 & 1,2,3 \\
NIR SpT & L1.5 + L4.5 & 3 \\
d$_{est}$  & 19$\pm$2 pc & 3 \\
$\rho$ & 1$\farcs$174$\pm$0$\farcs$016 & 3 \\
 & 22$\pm$2 AU & 3 \\
$\phi$ & 152$\fdg$9$\pm$0$\fdg$7 & 3 \\
$V_{tan}$ & 66$\pm$7~{\kms} & 3 \\
$V_{rad}$ & $-70{\pm}18$~{\kms} & 3 \\
($U,V,W$) & ($50,30,-20$)~{\kms} & 3 \\
$R_{ESO}$\tablenotemark{c} & 19.41 mag & 2 \\
$I_{N}$\tablenotemark{c} & 17.04 mag & 2 \\
$J$\tablenotemark{c} & 13.23$\pm$0.03 mag & 1 \\
$H$\tablenotemark{c} & 12.36$\pm$0.03 mag & 1 \\
$K_s$\tablenotemark{c} & 11.89$\pm$0.03 mag & 1 \\
${\Delta}J$ & 1.15$\pm$0.08 mag &  3 \\
${\Delta}H$ & 0.97$\pm$0.06 mag  &  3 \\
${\Delta}K$ & 0.95$\pm$0.03 mag &  3 \\
M$_{tot}$\tablenotemark{d} & 0.14--0.16~M$_{\sun}$ &  3,4 \\
Period\tablenotemark{d} & $\sim$400 yr & 3,4 \\
\enddata
\tablenotetext{a}{Equinox J2000 coordinates at epoch 1999 May 19 from 2MASS.}
\tablenotetext{b}{Estimates of systematic uncertainty based on PSF fitting simulations; see $\S$~2.1.}
\tablenotetext{c}{Photometry for combined (unresolved) system.}
\tablenotetext{d}{Assuming an age of 1-10~Gyr.}
\tablerefs{(1) 2MASS \citep{skr06}; (2) SSS \citep{ham01a,ham01b,ham01c}; (3) This paper;
(4) \citet{bur97}.}
\end{deluxetable}

\begin{deluxetable}{lcccccccc}
\tabletypesize{\footnotesize}
\tablecaption{Properties of the {\name}AB Components \label{tab_component}}
\tablewidth{0pt}
\tablehead{
 & & & \multicolumn{3}{c}{2MASS} & \multicolumn{3}{c}{Estimated Mass\tablenotemark{b}} \\
\cline{4-6} \cline{7-9}
\colhead{Component} &
\colhead{NIR SpT} &
\colhead{\teff\tablenotemark{a}} &
\colhead{$J$} &
\colhead{$H$} &
\colhead{$K_s$} &
\colhead{1 Gyr}  &
\colhead{5 Gyr}  &
\colhead{10 Gyr} \\
\colhead{} &
\colhead{} &
\colhead{(K)} &
\colhead{(mag)} &
\colhead{(mag)} &
\colhead{(mag)} &
\colhead{(M$_{\sun}$)} &
\colhead{(M$_{\sun}$)} &
\colhead{(M$_{\sun}$)} \\
}
\startdata
{\namesh}A & L1.5  & 2200 & 13.55$\pm$0.04 & 12.73$\pm$0.03 & 12.27$\pm$0.03 & 0.075 & 0.082 & 0.082  \\
{\namesh}B & L4.5  & 1740 & 14.70$\pm$0.07 & 13.70$\pm$0.05 & 13.22$\pm$0.04 & 0.064 & 0.077 & 0.078  \\
\enddata
\tablenotetext{a}{{\teff} estimate from \citet{vrb04}.}
\tablenotetext{b}{Based on the solar metallicity models of \citet{bur97}.}
\end{deluxetable}

\begin{deluxetable}{lcccccc}
\tabletypesize{\small}
\tablecaption{Near-infrared Spectral Classification Indices \label{tab_classify}}
\tablewidth{0pt}
\tablehead{
 & \multicolumn{2}{c}{{\namesh}A} & \multicolumn{2}{c}{{\namesh}B} & \multicolumn{2}{c}{{\namesh}AB (XD)} \\
\cline{2-3} \cline{4-5} \cline{6-7}
\colhead{Index} &
\colhead{Value} &
\colhead{SpT} &
\colhead{Value} &
\colhead{SpT} &
\colhead{Value} &
\colhead{SpT} \\
}
\startdata
\multicolumn{7}{c}{\citet{rei01a}} \\
\hline
{\water}-A & 0.686  &  L1.5 & 0.597  & L4  & 0.656 & L2.5 \\
{\water}-B & 0.749  &  L2 & 0.639 & L5 & 0.718 & L3 \\
$K1$\tablenotemark{a} & 0.178  &  L1 & 0.297  & L3.5 & 0.219 & L2 \\
\hline
\multicolumn{7}{c}{\citet{tes01}} \\
\hline
s{\water}$^{J}$ &  0.165 &  L2.5 & 0.259 & L4 & 0.229 & L3.5 \\
s{\water}$^{H1}$ & 0.327  &  L2 & 0.510 & L4 & 0.343 & L2  \\
s{\water}$^{H2}$ & 0.443  &  L2 & 0.538 & L4 & 0.517 & L4 \\
s{\water}$^{K}$ &  0.197 &  L0.5 & 0.332 & L4 & 0.255 & L2 \\
\hline
\multicolumn{7}{c}{\citet{me02a}} \\
\hline
{\water}-A & 0.878  &  L1.5 & 0.806  & L5 & 0.835 & L3.5 \\
{\water}-B & 0.857  &  L1.5 & 0.790  & L5 & 0.836 & L2.5 \\
{\water}-C & 0.868  &  L1 & 0.782  & L6.5 & 0.861 & L1.5 \\
{\meth}-C & 0.991  &  \nodata & 0.954  & L4.5  & 0.990 & \nodata \\
\hline
\multicolumn{7}{c}{\citet{geb02}} \\
\hline
{\water} 1.5$\micron$ & 1.333  &  L1.5 & 1.592  & L5.5 & 1.407 & L2 \\
{\meth} 2.2$\micron$ & 0.973 &  \nodata\tablenotemark{b} & 1.013 & L5 & 0.975 & \nodata\tablenotemark{b} \\
\hline
Mean NIR SpT & \multicolumn{2}{c}{L1.5$\pm$0.6} & \multicolumn{2}{c}{L4.5$\pm$0.8} & \multicolumn{2}{c}{L2.5$\pm$0.8} \\
\enddata
\tablenotetext{a}{Index defined in \citet{tok99}.}
\tablenotetext{b}{Index/SpT relation not defined for SpT $<$ L3 \citep{me02a,geb02}.}
\end{deluxetable}

\begin{deluxetable}{lcccccll}
\tabletypesize{\scriptsize}
\tablecaption{VLM Binaries Wider than 0$\farcs$5 \label{tab_widebinary}}
\tablewidth{0pt}
\tablehead{
\colhead{Name} &
\colhead{$\rho$} &
\colhead{d} &
\colhead{$R$\tablenotemark{a}} &
\colhead{$J$\tablenotemark{a}} &
\colhead{${\Delta}J$} &
\colhead{SpT} &
\colhead{References} \\
\colhead{}  &
\colhead{($\arcsec$)}  &
\colhead{(pc)} &
\colhead{(mag)} &
\colhead{(mag)} &
\colhead{(mag)} &
\colhead{} &
\colhead{} \\
}
\startdata
DENIS J020529.0-115925AB\tablenotemark{b,c} & 0$\farcs$51 & 12 & \nodata &  14.59 & 0.0 &  L5 + L7 & 1,2,3,4 \\
2MASS J04291842-3123568AB & 0$\farcs$53 & $\sim$11  & 16.2 &  10.89 & 1.2 &  M7.5 + [L1]\tablenotemark{d} & 5,6,7 \\
Gliese 337CD & 0$\farcs$53 & 21 & \nodata & 15.51 & 0.3 & L8 + [T:]\tablenotemark{d} & 8,9 \\
2MASS J23310161-0406193AB & 0$\farcs$57 & $\sim$26 & 18.9 &  12.94 & 2.8 &  M8 + [L7]\tablenotemark{d} & 4,10,11,12 \\
2MASS J09153413+0422045AB & 0$\farcs$73 & $\sim$15  & \nodata & 14.55 & 0.1 & L7 + [L7]\tablenotemark{d} & 7 \\
Eps Indi Bab & 0$\farcs$73 & 3.6 & 20.8 &  12.29 & 0.9 & T1 + T6 & 13,14 \\
2MASS J12073347-3932540AB & 0$\farcs$78 & $\sim$53 & 19.3  & 13.00 & 7.0  & M8.5 + L: & 15,16,17,18,19 \\
2MASS J17072343-0558249AB & 1$\farcs$01 & $\sim$15 & 18.9 &  12.05 & 1.3 & M9 + L3 & 7,15,20 \\
DENIS J220002.0-303832.9AB & 1$\farcs$09 & $\sim$35 & 19.2 & 13.44 & 0.3 & M9 + L0 & 21,22 \\
{\bf {\name}AB} & 1$\farcs$17 & $\sim$19 & 19.4 &  13.23 & 1.2 & L1.5 + L4.5 & 23,24 \\
SCR 1845-6537AB & 1$\farcs$17 & 3.9 & 16.3 &  9.54 & 3.7 & M8.5 + [T5.5]\tablenotemark{d} &  25,26,27,28 \\
2MASS J11011926-7732383AB\tablenotemark{c} & 1$\farcs$44 & $\sim$170 & 19.4 & 13.10 & 0.9 & M7 + M8 & 29 \\
2MASS J16233609-2402209AB\tablenotemark{e} & 1$\farcs$70  & $\sim$125 & \nodata & 14.62** & 0.7 & [M5:]\tablenotemark{d} + [M5:]\tablenotemark{d} & 30,31 \\
2MASS J16222521-2405139AB\tablenotemark{f} & 1$\farcs$94  & $\sim$125 & \nodata & 10.79** & 0.8 & M9 + M9.5 & 30,31,32,33 \\
DENIS J055146.0-443412.2AB & 2$\farcs$20 & $\sim$100  & \nodata  & 15.79 & 0.6 & M8.5 + L0  & 34 \\
\enddata
\tablenotetext{a}{$R$ and $J$ photometry for combined system from SSS and 2MASS, respectively; see also associated references.}
\tablenotetext{b}{Possible triple system \citep{bou05}.}
\tablenotetext{c}{Resolved optical spectroscopy has been obtained for this system.}
\tablenotetext{d}{Spectral type based on resolved photometry; resolved spectroscopy has
not yet been reported for this source.}
\tablenotetext{e}{a.k.a.\ Oph 16 \citep{all06}.}
\tablenotetext{e}{a.k.a.\ Oph 11 \citep{all06}.}
\tablerefs{(1) \citet{del97}; (2) \citet{koe99}; (3) \citet{leg01}; (4) \citet{bou03};
(5) \citet{cru03}; (6) \citet{sie05}; (7) \citet{rei06hst}; (8) \citet{wil01};
(9) \citet{me05gl337cd}; (10) \citet{giz00};  (11) \citet{giz03}; (12) \citet{clo03};
(13) \citet{sch03}; (14) \citet{mcc04}; (15) \citet{giz02};
(16) \citet{cha04}; (17) \citet{cha05}; (18) \citet{mam05}; (19) \citet{moh06};
(20) \citet{mce06}; (21) \citet{ken04}; (22) \citet{me06d2200}; (23) This paper;
(24) \citet{ken06};
(25) \citet{ham04}; (26) \citet{hen04}; (27) \citet{bil06}; 
(28) Henry et al.\ in preparation; (29) \citet{luh04}; 
(30) \citet{all06}; (31) \citet{clo06}; (32) \citet{jay06};
(33) \citet{all07}; (34) \citet{bil05}}
\end{deluxetable}


\begin{thebibliography}{}

\bibitem[Ackerman \& Marley(2001)]{ack01}Ackerman, A.\ S., \& Marley, M.\ S.
2001, \apj, 556, 872

\bibitem[Allard et al.(2001)]{all01}Allard, F., Hauschildt, P.\ H.,
Alexander, D.\ R., Tamanai, A., \&  Schweitzer, A. 2001, \apj, 556, 357

\bibitem[Allers et al.(2006)]{all06}Allers, K.\ N., Kessler-Silacci, J.\ E.,
Cieza, L.\ A., \& Jaffe, D.\ T. 2006, \apj, 644, 364

\bibitem[Allers et al.(2007)]{all07}Allers, K.\ N., et al. 2007, \apj, submitted

\bibitem[Bailer-Jones(2004)]{bai04} Bailer-Jones, C.\ A.\ L. 2004, \aap, 419, 703

\bibitem[Bate, Bonnell, \& Bromm(2003)]{bat03} Bate, M.\ R., Bonnell, I.\ A.,
\& Bromm 2003, \mnras, 339, 577

\bibitem[Becklin \& Zuckerman(1988)]{bec88}Becklin, E.\ E., \& Zuckerman,
B. 1988, Nature, 336, 656

\bibitem[Biller et al.(2006)]{bil06} Biller, B.\ A., Kasper, M., Close, L.\ M., Brandner, W., \& Kellner, S. 2006, \apj, 641, L141

\bibitem[Bill\'eres et al.(2005)]{bil05}Bill\'eres, M., Delfosse, X., Beuzit, J.-L.,
Forveille, T., Marchal, L., \& Mart{\'{\i}}n, E.\ L. 2005, \aap, 440, L55

\bibitem[Bouy et al.(2003)]{bou03} Bouy, H., Brandner, W., Mart{\'{\i}}n, E.\ L.,
Delfosse, X., Allard, F., \& Basri, G. 2003, \aj, 126, 1526

\bibitem[Bouy et al.(2004)]{bou04} Bouy, H., Brandner, W., Mart{\'{\i}}n, E.\ L.,
Delfosse, X., Allard, F., Baraffe, I., Forveille, T., \&
Demarco, R. 2004, \aap, 424, 213

\bibitem[Bouy et al.(2005)]{bou05} Bouy, H., Mart{\'{\i}}n, E.\ L., Brandner, W., 
\& Bouvier, J. 2005, \aj, 129, 511

\bibitem[Brandner et al.(2004)]{bra04} Brandner, W., Mar\'in, E.\ L., Bouy, H., K\"ohler, R., Delfosse X., Basri, G., \& Andersen, M. 2004, \aap, 428, 205

\bibitem[Burgasser(2004)]{me1626}Burgasser, A.\ J. 2004 \apj, 614, L73

\bibitem[Burgasser et al.(2006a)]{me06hst}Burgasser, A.\ J., Kirkpatrick, J.\ D.,
Cruz, K.\ L., Reid, I.\ N., Leggett, S.\ K.,  Liebert, Burrows, A., \& Brown, M.\ E. 2006a, \apj, in press

\bibitem[Burgasser, Kirkpatrick \& Lowrance(2005)]{me05gl337cd}Burgasser, A.\ J., Kirkpatrick, J.\ D., \& Lowrance, P.\ J. 2005, \aj, 129, 2849

\bibitem[Burgasser et al.(2003)]{me03b}Burgasser, A.\ J., Kirkpatrick,
J.\ D., Reid, I.\ N., Brown, M.\ E., Miskey, C.\ L., \& Gizis, J.\ E. 2003,
\apj, 586, 512

\bibitem[Burgasser \& McElwain(2006)]{me06d2200}Burgasser, A.\ J., \& McElwain, M.\ W. 2006, \aj, 131, 1007

\bibitem[Burgasser et al.(2005)]{me0423}Burgasser, A.\ J., Reid, I.\ N.,
Leggett, S.\ J., Kirkpatrick, J.\ D., Liebert, J., \& Burrows, A.
2005, \apj, 634, L177

\bibitem[Burgasser et al.(2006b)]{me06ppv}Burgasser, A.\ J., Reid, I.\ N.,
Siegler, N., Close, L.\ M., Allen, P., Lowrance, P.\ J., \& Gizis, J.\ E.
2006b, in Planets and Protostars V, eds.\ B.\ Reipurth, D.\ Jewitt and K.\ Keil (Univ.\ Arizona Press: Tucson), in press

\bibitem[Burgasser et al.(2002)]{me02a} Burgasser, A.\ J., et al. 2002, \apj, 564, 421

\bibitem[Burrows et al.(2001)]{bur01}Burrows, A., Hubbard, W.\ B., Lunine,
J.\ I., \& Liebert, J. 2001, Rev.\ of Modern Physics, 73, 719

\bibitem[Burrows \& Sharp(1999)]{bur99}Burrows, A., \& Sharp, C.\ M. 1999, \apj,
512, 843

\bibitem[Burrows et al.(1997)]{bur97}Burrows, A., et al.\ 1997, \apj, 491, 856

\bibitem[Cannon(1984)]{can84}Cannon, R.\ D. 1984, in Astronomy with with Schmidt-Type Telescopes, Proc.\ IAU Coll.\ 78, ed.\ M.\ Cappaccioli 
(Dordrecht: Reidel), p.\ 25

\bibitem[Chabrier et al.(2000)]{cha00}Chabrier, G., Baraffe, I., Allard, F., \&
Hauschildt, P. 2000, \apj, 542, 464

\bibitem[Chauvin et al.(2004)]{cha04}Chauvin, G., Lagrange, A.-M., Dumas, C., Zuckerman, B., Mouillet, D.,
Song, I., Beuzit, J.-L., \& Lowrance, P. 2004, \aap, 425, L29

\bibitem[Chauvin et al.(2005)]{cha05} ---. 2005, \aap, 438, L25

\bibitem[Chiu et al.(2006)]{chi06} Chiu, K., Fan, X., Leggett, S.\ K.,
Golimowski, D.\ A., Zheng, W., Geballe, T.\ R., Schneider, D.\ P., \&
Brinkmann, J. 2006, \apj, in press

\bibitem[Close et al.(2003)]{clo03}Close, L.\ M., Siegler, N., Freed, M.,
\& Biller, B. 2003, \apj, 587, 407

\bibitem[Close et al.(2006)]{clo06}Close, L.\ M., et al. 2006, \apj, submitted

\bibitem[Cohen, Wheaton \& Megeath(2003)]{coh03}Cohen, M.\ Wheaton, W.\ A., \& Megeath, S.\ T. 2003, \aj, 126, 1090


\bibitem[Corbally, Gray \& Garrison(1994)]{cor94} Corbally, C.\ J., Gray R.\ O.,
\& Garrison, R.\ F. 1994, The MK process at 50 years. A Powerful Tool for Astrophysical Insight
(San Francisco: Astronomical Society of the Pacific)

\bibitem[Cruz et al.(2004)]{cru04} Cruz, K.\ L., Burgasser, A.\ J., Reid, I.\ N., \&
Liebert, J. \apj, 2004, 604, L61

\bibitem[Cruz et al.(2003)]{cru03} Cruz, K.\ L., Reid, I.\ N., Liebert, J., Kirkpatrick, J.\ D.,
\& Lowrance, P.\ J. 2003, AJ, 126, 2421

\bibitem[Cushing et al.(2003)]{cus03}Cushing, M.\ C., Rayner, J.\ T., 
Davis, S.\ P., \& Vacca, W.\ D. 2003, \apj, 582, 1066

\bibitem[Cushing, Vacca, \& Rayner(2004)]{cus04} Cushing, M.\ C., Vacca, W.\ D., \& Rayner, J.\ T.
2004, PASP, 116, 362

\bibitem[Cushing, Rayner \& Vacca(2005)]{cus05} ---. 2005, \apj, 623, 1115

\bibitem[Cutri et al.(2003)]{cut03}Cutri, R.\ M., et al. 2003, Explanatory Supplement
to the 2MASS All Sky Data Release,
\url{http://www.ipac.caltech.edu/2mass/releases/allsky/doc/explsup.html}

\bibitem[Dahn et al.(2002)]{dah02}Dahn, C.\ C., et al. 2002, \aj,
124, 1170

\bibitem[Deacon, Hambly \& Cooke(2005)]{dea05} Deacon, N.\ R., Hambly, N.\ C., 
\& Cooke, J.\ A. 2005, \aap, 435, 363

\bibitem[Delfosse et al.(1997)]{del97}Delfosse, X., et al. 1997, \aap,
327, L25

\bibitem[Dehnen \& Binney(1998)]{deh98}Dehnen, W., \& Binney, J.\ J. 1998, \mnras,
298, 387


\bibitem[Filippenko(1982)]{fil82}Filippenko, A.\ V. 1982, \pasp, 94, 71

\bibitem[Geballe et al.(2002)]{geb02}Geballe, T.\ R., et al. 2002, \apj, 564, 466

\bibitem[Gizis(2002)]{giz02}Gizis, J.\ E. 2002, \apj, 575, 484

\bibitem[Gizis et al.(2000)]{giz00}Gizis, J.\ E., Monet, D.\ G., Reid, I.\ N.,
Kirkpatrick, J.\ D., Liebert, J., \& Williams, R. 2000, \aj, 120, 1085

\bibitem[Gizis et al.(2003)]{giz03}Gizis, J.\ E., Reid, I.\ N., Knapp, G.\ R., Liebert, J.,
Kirkpatrick, J.\ D., Koerner, D.\ W., \& Burgasser, A.\ J. 2003, \aj, 125, 3302

\bibitem[Gizis, Shipman \& Harvin(2005)]{giz05a}Gizis, J., Shipman, H., \& Harvin, J. 2005, \apj, 630, 89

\bibitem[Goldman et al.(1999)]{gol99}Goldman, B., et al. 1999, \aap, 351,
L5

\bibitem[Golimowski et al.(2004)]{gol04} Golimowski, D.\ A., et al. 2004, \aj, 127, 3516

\bibitem[Goto et al.(2002)]{got02}Goto, M., et al. 2002, 567, L59

\bibitem[Hambly et al.(2001a)]{ham01a}Hambly, N.\ C., Davenhall, A.\ C.,
Irwin, M.\ J., \& MacGillivray, H.\ T. 2001a, MNRAS 326, 1315

\bibitem[Hambly et al.(2004)]{ham04}Hambly, N.\ C., Henry, T.\ J., Subasavage, J.\ P., Brown, M.\ A., \& Jao, W.-C. 2004, \aj, 128, 437

\bibitem[Hambly et al.(2001b)]{ham01b}Hambly, N.\ C., Irwin, M.\ J., \& MacGillivray, H.\ T.
2001b, MNRAS 326, 1295

\bibitem[Hambly et al.(2001c)]{ham01c}Hambly, N.\ C., MacGillivray, H.\ T., Read, M.\ A.,
et al. 2001c, MNRAS 326, 1279

\bibitem[Harley \& Dawe(1981)]{har81}Hartley, M., \& Dawe, J.\ A. 1981, \pasa, 4, 251

\bibitem[Hawley et al.(2002)]{haw02}Hawley, S.\ L. et al. 2002, \aj, 123, 3409

\bibitem[Hayes(1985)]{hay85}Hayes, D.\ S. 1985, in IAU Symposium 111, Calibration of Fundamental Stellar Quantities, ed. D.\ S. Hayes, L.\ E. Pasinetti \& A.\ G.\ D.\ Philip (Dordrecht: D.\ Reidel) p.\ 225

\bibitem[Henry et al.(2004)]{hen04}Henry, T.\ J., Subasavage, J.\ P., Brown, M.\ A., Beaulieu, T.\ D., Jao, W.-C., \& Hambly, N.\ C. 2004, \aj, 128, 2460

\bibitem[Jayawardhana \& Ivanov(2006)]{jay06}Jayawardhana, R., \& Ivanov, V.\ D. 2006, Science, in press

\bibitem[Jones \& Tsuji(1997)]{jon97}Jones, H.\ R.\ A., \& Tsuji, T. 1997, \apj,
480, L39

\bibitem[Kirkpatrick(2005)]{kir05}Kirkpatrick, J.\ D. 2005, \araa, 43, 195

\bibitem[Kirkpatrick et al.(2006)]{kir06}Kirkpatrick, J.\ D., Barman, T.\ S., Burgasser, A.\ J., McGovern, M.\ R., McLean, I.\ S., Tinney, C.\ G., \&
Lowrance, P.\ J. 2006, \apj, 639, 1120

\bibitem[Kirkpatrick et al.(2000)]{kir00}Kirkpatrick, J.\ D., Reid, I.\ N.,
Liebert, J., Gizis, J.\ E., Burgasser, A.\ J., Monet, D.\ G., Dahn, C.\ C.,
Nelson, B., \& Williams, R.\ J. 2000, \aj, 120, 447

\bibitem[Kirkpatrick et al.(1999)]{kir99}Kirkpatrick, J.\ D., et al. 1999,
\apj, 519, 802

\bibitem[Kendall et al.(2004)]{ken04}Kendall, T.\ R., Delfosse, X., Mart{\'{\i}}n, E.\ L.,
\& Forveille, T. 2004, \aap, 416, L17

\bibitem[Kendall et al.(2006)]{ken06}Kendall, T.\ R., Jones, H.\ R.\ A., Pinfield, D.\ J., Porkorny, R.\ S., Folkes, S., Weights, D.\ Jenkins, J.\ S., 
\& Mauron, N. 2006, MNRAS, in press

\bibitem[Knapp et al.(2004)]{kna04}Knapp, G., et al. 2004, \apj, 127, 3553

\bibitem[Koerner et al.(1999)]{koe99}Koerner, D.\ W., Kirkpatrick, J.\ D.,
McElwain, M.\ W., \& Bonaventura, N.\ R. 1999, \apj, 526, L25

\bibitem[Kurucz \& Bell(1995)]{kur95}Kurucz, R.\ L., \& Bell, B. 1995, Atomic Line Data, Kurucz CD-ROM No. 23. (Cambridge: Smithsonian Astrophysical Observatory)

\bibitem[Lane et al.(2001)]{lan01} Lane, B.\ F., Zapatero Osorio, M.\ R.,
Britton, M.\ C., Mart{\'{\i}}n, E.\ L., \& Kulkarni, S.\ R. 2001, \apj,
560, 390

\bibitem[Leggett et al.(2001)]{leg01} Leggett, S.\ K., Allard, F.,
Geballe, T., Hauschildt, P.\ H., \& Schweitzer, A. 2001, \apj, 548, 908

\bibitem[Liu \& Leggett(2005)]{liu05}Liu, M.\ C., \& Leggett, S.\ K. 2005, \apj, 634, 616

\bibitem[Liu et al.(2006)]{liu06}Liu, M.\ C., Leggett, S.\ K., Golimowski,
D.\ A., Chiu, K., Fan, X., Geballe, T.\ R., Schneider, D.\ P., \& 
Brinkmann, J. 2006, \apj, in press

\bibitem[Lodieu et al.(2005)]{lod05}Lodieu, N., Scholz, R.-D., McCaughrean, M.\ J., Ibata, R., Irwin, M., \& Zinnecker, H. 2005, \aap, 440, 1061

\bibitem[Luhman(2004)]{luh04}Luhman, K.\ L. 2004, \apj, 614, 398

\bibitem[Mamajek(2005)]{mam05}Mamajek, E.\ E. 2005, \apj, 634, 1385

\bibitem[Mart{\'{\i}}n et al.(2006)]{mar06}Mart{\'{\i}}n, E.\ L., Brandner,
W., Bouy, H., Basri, G., Davis, J., Deshpande, R., Montgomery, M., \& King, I.
2006, \aap, in press

\bibitem[McCaughrean et al.(2004)]{mcc04}McCaughrean, M.\ J., 
Close, L.\ M., Scholz, R.-D., Lenzen, R., Biller, B., 
Brandner, W., Hartung, M., \& Lodieu, N. 2004, \aap, 413, 1029

\bibitem[McElwain \& Burgasser(2006)]{mce06} McElwain, M.\ W., \&
Burgasser, A.\ J. 2006, \aj, in press

\bibitem[McGovern(2005)]{mcg05}McGovern, M.\ R. 2005, Ph.D.\ Thesis, University of California Los Angeles

\bibitem[McGovern et al.(2004)]{mcg04}McGovern, M.\ R., Kirkpatrick, J.\ D., McLean, I.\ 
S., Burgasser, A.\ J., Prato, L., \& Lowrance, P.\ J. 2004, \apj
600, 1020

\bibitem[McLean et al.(2003)]{mcl03}McLean, I.\ S., McGovern, M.\ R., Burgasser, A.\ J.,
Kirkpatrick, J.\ D., Prato, L., \& Kim, S. 2003, \apj, 596, 561

\bibitem[Mohanty \& Basri(2003)]{moh03a}Mohanty, S., \& Basri, G. 2003, \apj, 583, 451

\bibitem[Mohanty et al.(2002)]{moh02}Mohanty, S., Basri, G., Shu, F., Allard, F., \& Chabrier, G. 2002,
\apj, 572, 469

\bibitem[Mohanty, Jayawardhana \& Barrado y Navascu{\'{e}}s(2003)]{moh03}Mohanty, S., Jayawardhana, R.,
\& Barrado y Navascu{\'{e}}s, D. 2003, \apj, 593, L109

\bibitem[Mohanty et al.(2006)]{moh06}Mohanty, S., Jayawardhana, R., Hu\'elamo, N, \& Mamajek, E. 2006, \apj, submitted

\bibitem[Morgan, Keenan, \& Kellman(1943)]{mor43} Morgan, W.\ W.,
Keenan, P.\ C., \& Kellman, E. 1943, An Atlas
of Stellar Spectra, with an Outline of Spectral Classification
(Chicago: Univ.\ Chicago Press)

\bibitem[Morgan et al.(1992)]{mor92}Morgan, D.\ H., Tritton, S.\ B., Savage, A., Hartley, M., \&
Cannon, R.\ D. 1992, in Digitised Optical Sky Surveys, ed. H.\ T.\ MacGillivray \&
E.\ B.\ Thomson (Dordrecht: Boston), p.\ 11


\bibitem[Mountain et al.(1985)]{mnt85}Mountain, C.\ M., Selby, M.\ J., 
Leggett, S.\ K., Blackwell, D.\ E., \& Petford, A.\ D. 1985, \aap, 151, 399

\bibitem[Mugrauer \& Neuh{\"{a}}user(2005)]{mug05}Mugrauer, M., \& Neuh{\"{a}}user, R. 2005, AN, 326, 701

\bibitem[Nakajima et al.(1995)]{nak95}Nakajima, T., Oppenheimer, B.\ R.,
Kulkarni, S.\ R.,
Golimowski, D.\ A., Matthews, K., \& Durrance, S.\ T. 1995, \nat, 378, 463

\bibitem[Nakajima et al.(2004)]{nak04} Nakajima, T., Tsuji, T., \&
Yanagisawa, K. 2004, \apj, 607, 499

\bibitem[Potter et al.(2002)]{pot02}Potter, D., Mart{\'{\i}}n, E.\ L.,
Cushing, M.\ C., Baudoz, P., Brandner, W., Guyon, O., \& Neuh\"{a}user, R.
2002, \apj, 567, L133

\bibitem[Rayner et al.(2003)]{ray03} Rayner, J.\ T., Toomey, D.\ W., Onaka, P.\ M., Denault, A.\ J.,
Stahlberger, W.\ E., Vacca, W.\ D., Cushing, M.\ C., \& Wang, S. 2003, PASP, 155, 362

\bibitem[Rebolo, Mart{\'{\i}}n, \& Magazzu(1992)]{reb92}Rebolo, R.,
Mart{\'{\i}}n, E.\ L., \& Magazzu, A. 1992, \apj, 389, L83

\bibitem[Rebolo et al.(1998)]{reb98}Rebolo, R., Zapatero Osorio, M.\ R.,
Madruga, S., B{\'{e}}jar, V.\ J.\ S., Arribas, S., \& Licandro, J.
1998, Science, 282, 1309

\bibitem[Reid et al.(2001a)]{rei01a}Reid, I.\ N., Burgasser, A.\ J.,
Cruz, K., Kirkpatrick, J.\ D., \& Gizis, J.\ E. 2001a, \aj, 121, 1710

\bibitem[Reid et al.(2001b)]{rei01}Reid, I.\ N., Gizis, J.\ E., Kirkpatrick,
J.\ D., \& Koerner, D. 2001b, \aj, 121, 489

\bibitem[Reid et al.(2006)]{rei06hst}Reid, I.\ N., Lewitus, E., Allen, P.\ R.,
Cruz, K.\ L., \& Burgasser, A.\ J. 2006, \aj, 132, 891

\bibitem[Reid et al.(1991)]{rei91}Reid, I.\ N., et al. 1991, \pasp, 103, 661

\bibitem[Reipurth \& Clarke(2001)]{rpt01}Reipurth, B., \& Clarke, C. 2001, \aj, 122, 432

\bibitem[Roe(2002)]{roe02} Roe, H.\ G. 2002, \pasp, 114, 450

\bibitem[Ruiz, Leggett, \& Allard(1997)]{rui97}Ruiz, M.\ T., Leggett,
S.\ K., \& Allard, F. 1997, \apj, 491, L107

\bibitem[Scholz et al.(2003)]{sch03}Scholz, R.-D.,  McCaughrean, M.\ J., Lodieu, N.,
\&  Kuhlbrodt, B. 2003, \aap, 398, L29

\bibitem[Scholz \& Meusinger(2002)]{sch02} Scholz, R.-D., \& Meusinger, H. 2002, \mnras, 336, L49

\bibitem[Schubert \& Walterscheid(2000)]{sch00}Schubert, G., \& 
Walterscheid, R.\ L. 2000, in Allen's Astrophysical Quantities, ed.\ A.\ N.\ 
Cox (4th ed.; New York: AIP), 239

\bibitem[Siegler et al.(2005)]{sie05}Siegler, N., Close, L.\ M., Cruz, K.\ L.,
Mart{\'{\i}}n, E.\ L., \& Reid, I. N. 2005, \apj, 621, 1023

\bibitem[Simons \& Tokunaga(2002)]{sim02}Simons, D.\ A., \& Tokunaga, A.\ T. 2002, \pasp, 114, 169

\bibitem[Skrutskie et al.(2006)]{skr06}Skrutskie, M.\ F., et al. 2006, \aj, 131, 1163

\bibitem[Stassun et al.(2006)]{sta06}Stassun, K., Mathieu, R.\ D., Vaz, L.\ P.\ R., 
Valenti, J.\ A., \& Gomez, Y. 2006, Nature, 440, 311

\bibitem[Stephens(2001)]{ste01}Stephens, D.\ C. 2001, Ph.D.\ Thesis, New Mexico State University

\bibitem[Stephens(2003)]{ste03}Stephens, D.\ C. 2003, in IAU Symposium 211, Brown Dwarfs, ed.\ E. Mart\'in (San Francisco: ASP), p.\ 355

\bibitem[Testi et al.(2001)]{tes01}Testi, L., et al. 2001, \apj, 522, L147

\bibitem[Tinney, Burgasser, \& Kirkpatrick(2003)]{tin03}Tinney, C.\ G., Burgasser, A.\ J.,
\& Kirkpatrick, J.\ D. 2003, \aj, 126, 975

\bibitem[Tokunaga \& Kobayashi(1999)]{tok99}Tokunaga, A.\ T., \&
Kobayashi, N. 1999, \aj, 117, 1010

\bibitem[Tokunaga, Simons \& Vacca(2002)]{tok02}Tokunaga, A.\ T., Simons, D.\ A.,
\& Vacca W.\ D. 2002, \pasp, 114, 180

\bibitem[Tsuji(2005)]{tsu05}Tsuji, T. 2005, \apj, 621, 1033

\bibitem[Tsuji, Ohnaka, \& Aoki(1996a)]{tsu96a}Tsuji, T., Ohnaka, K., \&
Aoki, W. 1996, \aap, 305, L1

\bibitem[Vacca et al.(2003)]{vac03}Vacca, W.\ D., Cushing, M.\ C., \& Rayner, J.\ T. 2003,
PASP, 155, 389

\bibitem[Vrba et al.(2004)]{vrb04}Vrba, F.\ J., et al. 2004, \aj, 127, 2948


\bibitem[West et al.(2004)]{wes04}West, A.\ A., et al. 2004, \aj, 128, 426

\bibitem[Wilson et al.(2001)]{wil01}Wilson, J.\ C., et al. 2001, \aj,
122, 1989

\bibitem[Zapatero Osorio et al.(2004)]{zap04}Zapatero Osorio, M.\ R., Lane, B.\ F., Pavlenko, Ya.,
Mart{\'{i}}n, E. L., Britton, M., \& Kulkarni, S. R. 2004, \apj, 615, 958

\end{thebibliography}
\end{document}